\documentclass[12pt]{article}

\usepackage{physics} 
\usepackage{url}
\usepackage{graphicx}
\usepackage{siunitx} 
\usepackage{enumerate} 
\usepackage{enumitem}
\usepackage{makecell}
\usepackage{pgfplots}
\usepackage{pgfplotstable}
\usepackage{tikz,pgfplots}
\usepackage{amssymb}
\usepackage{bbm}
\usepackage{amsmath}  %I added this so that you can use the align tool for equations!
\usepackage{cleveref}
\usepackage{titlesec}
\usepackage{wasysym} %This package allows you to put emojis in your paper!!!!
\usepackage{subcaption}
\usepackage{geometry}
\usepackage[toc,page]{appendix}
\usepackage{booktabs}
\usepackage{tabularx}
\usepackage{makecell}

\titleclass{\subsubsubsection}{straight}[\subsection]

\newcounter{subsubsubsection}[subsubsection]
\renewcommand\thesubsubsubsection{\thesubsubsection.\arabic{subsubsubsection}}
\titleformat{\subsubsubsection}
  {\normalfont\normalsize\bfseries}{\thesubsubsubsection}{1em}{}
\titlespacing*{\subsubsubsection}
{0pt}{3.25ex plus 1ex minus .2ex}{1.5ex plus .2ex}

\renewcommand\paragraph{\@startsection{paragraph}{5}{\z@}%
  {3.25ex \@plus1ex \@minus.2ex}%
  {-1em}%
  {\normalfont\normalsize\bfseries}}
\renewcommand\subparagraph{\@startsection{subparagraph}{6}{\parindent}%
  {3.25ex \@plus1ex \@minus .2ex}%
  {-1em}%
  {\normalfont\normalsize\bfseries}}
\def\toclevel@subsubsubsection{4}
\def\toclevel@paragraph{5}
\def\toclevel@paragraph{6}
\def\l@subsubsubsection{\@dottedtocline{4}{7em}{4em}}
\def\l@paragraph{\@dottedtocline{5}{10em}{5em}}
\def\l@subparagraph{\@dottedtocline{6}{14em}{6em}}
\makeatother

\pgfplotsset{compat=1.14}
\usepackage{authblk}

\author[1,2,*,$\diamondsuit$]{Joel Dyer}
\author[1,2,$\dagger$,$\diamondsuit$]{Blas Kolic}
\affil[1]{{Institute for New Economic Thinking at the Oxford Martin School, University of Oxford}}%, Oxford OX2 6ED, UK}}
\affil[2]{{Mathematical Institute, University of Oxford}}%, Oxford OX2 6GG, UK}}
\affil[$\dagger$]{\textit{blas.kolic@maths.ox.ac.uk}}
\affil[*]{\textit{joel.dyer@maths.ox.ac.uk}}
\affil[$\diamondsuit$]{Equal contribution}

\begin{document}

% \title{Weber-Fechner's law explains the perception of Covid-19 deaths in Twitter}
% \title{The public's risk perception and emotion in the Covid-19 pandemic}
% \title{ Modeling the public perception and emotion in the Covid-19 pandemic through psychophysical numbing: a Twitter data-driven approach  }
\title{ Public risk perception and emotion on Twitter during the Covid-19 pandemic }

\date{\today}

\maketitle
\tableofcontents

\newpage

\begin{abstract}
    Successful navigation of the Covid-19 pandemic is predicated on public cooperation with safety measures and appropriate perception of risk, in which emotion and attention play important roles. Signatures of public emotion and attention are present in social media data, thus natural language analysis of this text enables near-to-real-time monitoring of indicators of public risk perception. We compare key epidemiological indicators of the progression of the pandemic with indicators of the public perception of the pandemic constructed from $\sim 20$ million unique Covid-19-related tweets from 12 countries posted between 10th March -- 14th June 2020. We find evidence of psychophysical numbing: Twitter users increasingly fixate on mortality, but in a decreasingly emotional and increasingly analytic tone. {  Semantic network analysis based on word co-occurrences reveals changes in the emotional framing of Covid-19 casualties that are consistent with this hypothesis.} {  We also find that the average attention afforded to national Covid-19 mortality rates is modelled accurately with the Weber-Fechner and power law functions of sensory perception}. Our parameter estimates for these models are consistent with estimates from psychological experiments, and indicate that users in this dataset exhibit differential sensitivity by country to the national Covid-19 death rates. Our work illustrates the potential utility of social media for monitoring public risk perception and guiding public communication during crisis scenarios.
\end{abstract}


\section{Introduction}

The Covid-19 pandemic has brought about widespread disruption to human life. In many countries, public gatherings have been broadly forbidden, mass restrictions on human movement have been introduced, and entire industries have been paralysed in attempting to lower the peak stress on healthcare systems \cite{SI}. However, the degree to which these restrictions have been enforced by law has varied over time and by location, and their success in mitigating public health risks depends on the extent of cooperation on the part of the public.

A key determinant of the public's behaviour and their cooperation with state-imposed social restrictions is the public's emotional response to, and their perception of the the risk presented by, the pandemic. However, the evolution of emotions and risk perception in response to disasters is not well-understood, and there is a need for more longitudinal data on such responses with which this understanding can be improved \cite{BurnsSlovic2012}. Our goal is thus to contribute to bettering this understanding, and we do so by exploring the empirical relationships present between the progression of the Covid-19 pandemic and the public's perception of the risk posed by the pandemic. 

We explain our findings in terms of the existing body of literature surrounding public perception of risk, disasters, and human suffering in cognitive psychology. In particular, we draw from psychophysics, the field that studies the relationship between stimulus and subjective sensation and perception \cite{Geisch1997}. The search for psychophysical ``laws'' of perception has existed since at least the mid-19th Century with the proposing of the Weber-Fechner law \cite{fechner1966elements}, which posits that the smallest perceptible change ${\rm d}s$ in a physical stimulus of magnitude $s$ is proportional to $s$. Thus, the perceived magnitude $p$ of such stimuli follows
\begin{equation}\label{eq:WFL_}
    {\rm d}p \propto \frac{{\rm d}s}{s}.
\end{equation}
In the continuum limit, this implies that $p$ grows logarithmically with the physical magnitude $s$ of the stimulus.
% \begin{equation}
%     p = k \log{\frac{s}{s_0}}.
% \end{equation}
More recently, empirical studies by S. S. Stevens \cite{Stevens1975} supported, instead, a power law relationship between human perception of a stimulus and the physical magnitude of the stimulus:
\begin{equation}
    p \propto s^{\beta}.
\end{equation}
Summers {\it et al.} \cite{Summers1994} extended this concept to human sensitivity to war death statistics and found that a power law with exponent $\beta = 0.32$ best fit the data. A number of further studies have corroborated the extension of these psychophysical laws describing the subjective perception of physical magnitudes to the subjective evaluations of human fatalities \cite{Slovic2010, Fether1997, Fried1999}. In all of these, perception is a concave function of the stimulus, meaning that the larger the stimulus magnitude, the more it has to change in absolute terms to be equally noticeable. Thus, perception is considered relative rather than absolute, implying that our judgments are comparative in nature. This observation has been shown to account for deviations from rationality in economic decision-making \cite{weber2004perception}.

These proposed psychophysical laws of human perception present an opportunity for monitoring a population's response to a disaster scenario such as the Covid-19 pandemic. By evaluating the goodness of fit of these models to data on the perception of the progression of the pandemic, and determining the parameter values of such fits, we can describe the sensitivity of populations to the state of such crises, with important implications for risk communication and disaster management.

To this end, we make use of a massive Twitter dataset consisting of user-posted textual data to study the public's emotional and perceptual responses to the current public health crisis. Twitter provides convenient access to the conversation amongst members of the public across the globe on a plethora of topics, and many authors are studying several aspects of the public's response to the pandemic with it. %\cite{ferrara2020covid, alshaabi2020world, sha2020dynamic, Stella}. 
Twitter is a particularly appropriate tool under conditions of physical distancing requirements and furlough schemes, where online communication has become more than ever a central feature of everyday life. Moreover, results from psycholinguistics and advances in natural language processing techniques enable the extraction of psychologically meaningful attributes {  and the reconstruction of cognitive structures (e.g. semantic networks)} from textual data. With this dataset, our general approach is to offer a quantitative, spatiotemporal comparison between indicators of the state of the pandemic and the topics and psychologically meaningful linguistic features present in the discussion surrounding Covid-19 on social media on a country-by-country basis, for a selection of countries.

\subsection{Related work}

Our work is novel in that, to our knowledge, it is the first to use a large social media dataset spanning multiple countries to model the perceptual response of countries' citizens to the pandemic in the context of risk perception. To date, empirical validation of the aforementioned psychophysical laws has largely taken place in controlled laboratory settings, in which decisions, actions, and scenarios are artificial or hypothetical. Our work thus contributes to the body of literature surrounding risk perception by investigating these laws in a naturalistic setting. 

However, there have been numerous authors using social media to analyse the public response to the Covid-19 pandemic. This includes work that has focused on the psychological burden of the social restrictions. For instance, Stella et al. \cite{Stella} use the circumplex model of affect \cite{CircModAff} and the NRC lexicon \cite{NRCLex} to give a descriptive analysis of the public mood in Italy from a Twitter dataset collected during the week following the introduction of lockdown measures. In addition, Venigalla {\it et al} \cite{Venigalla} has developed a web portal for categorising tweets by emotion in order to track mood in India on a daily basis.

Others have instead focused on negative emotions, as in the work of Schild et al. \cite{Schild}, where they study the rise of hate speech and sinophobia as a result of the outbreaks. More specifically on perception, Dryhust et al. \cite{Dryhurst} measured the perceived risk of the Covid-19 pandemic by conducting surveys at a global scale ($n \sim 6000$) and compared countries, finding that factors such as individualistic and pro-social values and trust in government and science were significant predictors of risk perception. de Bruin and Bennett \cite{deBruin} perform similar work in the United States. The closest work we have been able to find to our own are those of Barrios and Hochberg \cite{Barrios} {   and Aiello et al. \cite{aiello2020epidemic}, where both research pieces focus on the current pandemic using data from the United States. In the former, they} combine internet search data with daily travel data to show that regions in the United States with a greater proportion of Trump voters exhibit behaviours that are consistent with a lower perceived risk during the Covid-19 pandemic. {  In the latter, they assess the epidemic psychology using Covid-19 Twitter data in the United States according to several linguistic features present in the tweets. They identify three psychological phases consistent with the refusal-suspended reality-acceptance stages of grief}. Despite the above, we have been unable to find work that combines large-scale social media data with linguistic analysis to offer a spatiotemporal, quantitative analysis of emotion and risk perception during the Covid-19 pandemic across multiple countries.

% CITE OTHER TWITTER STUDIES AROUND COVID:
% - H. Sha: Dynamic topic modeling of the COVID-19 Twitter narrative among U.S. governors and cabinet executives. Hawkes model and granger causality network of twitter data
% -  DONE L. Schild go eat a bat
% - DONE M. Stella: #lockdown: network-enhanced emotional profiling at the times of COVID-19. They develop thermometer to gauge emotional status.
% - T. Alshaabi: How the world’s collective attention is being paid to a pandemic: COVID-19 related 1-gram time series for 24 languages on Twitter. 
% - E. Ferrara: #COVID-19 ON TWITTER : BOTS , CONSPIRACIES, AND SOCIAL MEDIA ACTIVISM
% - C. Ordun: Exploratory Analysis of Covid-19 Tweets using Topic Modeling, UMAP, and DiGraphs
% - DONE A. Sri Manasa Venigalla: Mood of India During Covid-19 - An Interactive Web Portal Based on Emotion Analysis of Twitter Data

% CITE OTHER RISK & PERCEPTION COVID WORK:
% - T. Fetzter: Mental well-being at the onset of Covid-19
% - DONE S. Dryhurst: Risk perceptions of COVID-19 around the world. Poll with 6k samples in 10 countries
% - J. Van Bavel: Using social and behavioural science to support COVID-19 pandemic response (Nature Human behaviour)
% - 

Beyond the Covid-19 pandemic, our work is related to a small but growing body of literature on the use of data science in understanding human emotion and risk perception. In such work, natural language analysis has succeeded in supporting established linguistic theories such as the importance of the distribution of words in a vocabulary as a proxy for knowledge \cite{harris1954distributional}, and regarding the relation between the uncertainty of events and the emotional response to their outcome \cite{feather1963effect, verinis1968discrepancy}. For instance, using textual data from Twitter, Bhatia found that unexpected events elicit higher affective responses than those which are expected \cite{bhatia2019affective}. In another instance, the same author conducted experiments with $300$ participants and predicted the perceived risk of several risk sources using a vector-space representation of natural language, concluding that the word distribution of language successfully captures human perception of risk \cite{bhatia2019predicting}. Similar work has been conducted by Jaidka {\it et al}. \cite{Jaidka2020} in the area of monitoring public well-being, in which they compare word-based and data-driven methods for predicting ground-truth survey results for subjective well-being of US citizens on a county-level basis using a 1.5 billion Tweet dataset constructed from 2009 to 2015.

The remainder of this paper is laid out as follows. In Section \ref{Data}, we present the data set used in the subsequent analysis. In Section \ref{Method}, we provide further details on the approach followed to explore the relationships between indicators of the state of the pandemic and the public's perception of the pandemic, and discuss possible explanations for our observations by drawing on psychological literature. In Section \ref{Conc}, we summarise and offer concluding remarks, along with a discussion of the limitations of the current work and suggestions for avenues of future work.

\section{Data}\label{Data}

\subsection{Twitter dataset}\label{TwitterData}

In the following analysis, we make use of the set of tweets gathered by J. Banda et. al \cite{banda_juan_m_2020_3757272}, which are obtained and mantained using the Twitter free Stream API\footnote{The free Stream API randomly samples around $1\%$ of the total tweets for the given queries}. At the time of writing, this data set consists of $\sim80$ million \textit{original} tweets spanning from March 11, 2020 to June 14, 2020. {  By original we mean that we do not consider retweets,  which is standard for natural language processing \cite{go2009twitter, banda_juan_m_2020_3757272}}. Data is collected according to the following query filters\footnote{A number of publicly available Twitter datasets have emerged in relation to the pandemic. We chose to work with this dataset since it used the most generic query terms among all the publicly available datasets we considered, and we wanted the least amount of bias possible for our analysis.}: ``COVID19'', ``CoronavirusPandemic'', ``COVID-19'', ``2019nCoV'', ``CoronaOutbreak'', ``coronavirus'', ``WuhanVirus'', ``covid19'', ``coronaviruspandemic'', ``covid-19'', ``2019ncov'', ``coronaoutbreak'', ``wuhanvirus''. 

For our analysis, we consider only the English and Spanish tweets with a non-empty self-reported location field. We process every self-reported location using OpenStreetMaps \cite{OpenStreetMap} and remove non-sensical locations (e.g. ``Mars'', ``Everywhere'', ``Planet Earth''). This allows us to group the remaining tweets by country and proceed with our analysis on a country-by-country basis. To assure the statistical significance of our analysis, we keep the countries with the highest number of tweets for each language, resulting in a geolocated Twitter dataset of $\sim20$ million {  original} tweets posted by $\sim 4$ million users on $12$ different countries, which we summarise in Table \ref{table:twitter_data}.

%TABLE
\begin{table}%[h!]
\centering
% \resizebox{\textwidth}{!}{%
\begin{tabular}{l|c|c|c}
\toprule
{}                      & \textbf{Language} & \textbf{Number of tweets} & \textbf{Unique users} \\
\midrule
\textbf{Argentina}      & Spanish            & 846,706    & 194,818   \\ 
\textbf{Australia}      & English            & 701,072    & 97,027    \\ 
\textbf{Canada}         & English            & 1,209,712  & 195,507   \\ 
\textbf{Chile}          & Spanish            & 342,013    & 60,235    \\ 
\textbf{Colombia}       & Spanish            & 466,477    & 103,845   \\ 
\textbf{India}         & English            & 1,806,685  & 344,894   \\ 
\textbf{Mexico}         & Spanish            & 1,133,350  & 187,064   \\ 
\textbf{Nigeria}        & English            & 754,152    & 133,797   \\ 
\textbf{South Africa}   & English            & 354,613    & 78,447    \\ 
\textbf{Spain}          & Spanish            & 1,697,049  & 274,010   \\ 
\textbf{United Kingdom} & English            & 3,490,703  & 631,017   \\ 
\textbf{United States}  & English \& Spanish & 6,297,720  & 1,397,410 \\ \hline
\textbf{Total}          &                    & 19,072,850 & 3,699,071 \\
\bottomrule
\end{tabular}%
% }
\caption{Per-country summary of the Twitter dataset constructed from the repository maintained by Banda et al. \cite{banda_juan_m_2020_3757272}. {  All tweets are original, i.e. all retweets are removed.}
}
\label{table:twitter_data}
\end{table}

\subsection{Epidemiological data}\label{EpData}

We measure the progression of the pandemic with the number of Covid-19 confirmed cases and deaths for all the countries in our analysis. The data was made publicly available by Our World in Data repository \cite{owidcoronavirus}. In particular, we take the daily Covid-19 cases and deaths, both in linear and logarithmic scale, since these are four epidemiological indicators that are most frequently used to summarise the state of the pandemic, and are therefore frequently encountered by the public.

\section{Analysing the public's perception of the pandemic}\label{Method}

In this section, we study the public's perception of the pandemic on a country-by-country basis, using the countries with the highest number of tweets in the observation period (see Table \ref{table:twitter_data}). We do this on a country-by-country basis since the pandemic has often evoked nation-level responses, making nation-level analysis the most natural geographic scale. Our broad approach is to inspect and compare the linguistic features of the tweets released by users in the Twitter dataset described in Section \ref{TwitterData} with the epidemiological data described in Section \ref{EpData}.

\subsection{Defining perception from linguistic inquiry}\label{PercLing}

Our goal is to explore the public's perception of the pandemic. To do this, we analyse the linguistic features present in the textual data generated by Twitter users, and map these features to psychologically meaningful categories that are indicative of the Twitter users' perception. Here, we are assuming that the words used by these Twitter users are indicative of their internal cognitive and emotional states \cite{tausczik2010psychological}, which is supported in \cite{bhatia2019predicting} where they predict the perception of risk using text data. Thus, we quantify the linguistic content of each tweet using the Linguistic Inquiry and Word Count (LIWC) program \cite{PennLIWC2015}. LIWC has been widely adopted in several text data analyses, and it has proven successful in applications ranging from measuring the perception of emotions \cite{yin2014anxious} to predicting the German federal elections using Twitter \cite{tumasjan2010predicting}. {  Moreover, it has recently been used to successfully identify the early-epidemic psychological stages of grief in the current pandemic \cite{aiello2020epidemic}}.

LIWC operates as text analysis program that reports the number of words in a document belonging to a set of predefined linguistically and psychologically meaningful categories\footnote{For the English-language tweets, we make use of the 2015 English dictionary. For the Spanish-language tweets, the most recent dictionary is the 2007 edition, which has fewer categories than the 2015 English dictionary.} \cite{tausczik2010psychological}. For our purposes, a document is a tweet $d_i^t$ posted on date $t$ and from a user based in country $i$. LIWC represents documents as an unordered set of words, and a LIWC category $l$ is similarly a set of words associated with concept $l$. For a given document $d_i^t$, the {\it linguistic score} $p^l$ for category $l$ is the percentage of words in $d_i^t$ that belong to $l$:
\begin{equation}
    p^l(d_i^t) = \frac{ \vert{d_i^t \cap l}\vert }{ \vert{d_i^t}\vert } \cdot 100.
    \label{eq:liwc_score_individual}
\end{equation}

There are many such categories $l$, including Family, Work, and Motion. We capitalise such category titles, and use the titles to refer to either the set of words associated with that category or to refer to the category itself. Linguistic scores from Eq. (\ref{eq:liwc_score_individual}) for individual tweets will be noisy, as they are short documents.  Moreover, we are interested in the average response of the population of a country.  For this reason, we group the tweets by country $i$ and by date $t$, and denote these sets of tweets as $D^t_i = \{\ d_{i'}^{t'}\ |\ i' = i,\ t' = t\ \}$. We then compute the {\it National Linguistic Score} (NLS) for category $l$ as the average of the linguistic scores over documents in $D_i^t$ relative to an empirically observed Twitter base rate $p^l_B$: 
\begin{equation}
    p^l_i(t) = \frac{100}{\lvert D^t_i \rvert} \sum_{d \in D^t_i} \frac{p^l(d) - p^l_B}{p^l_B}.
    \label{eq:liwc_score}
\end{equation}

The base rates $p_B^l$ for the use of words on Twitter associated with category $l$ are given in \cite{PennLIWC2015}. Using Eq. (\ref{eq:liwc_score}) for all the selected linguistic categories, we construct multidimensional country-level time series that represent the evolution of the public perception of the pandemic, similar to the linguistic profiles introduced by Tumasjan {\it et al.} \cite{tumasjan2010predicting}. {  These perception dynamics are influenced by each user in our dataset, which may include bots and institutional or public relations accounts. We discuss the possible implications of this aspect of our data in Section \ref{Conc}.}

\begin{figure}%[h!]
    \centering
    \includegraphics[width=1\linewidth]{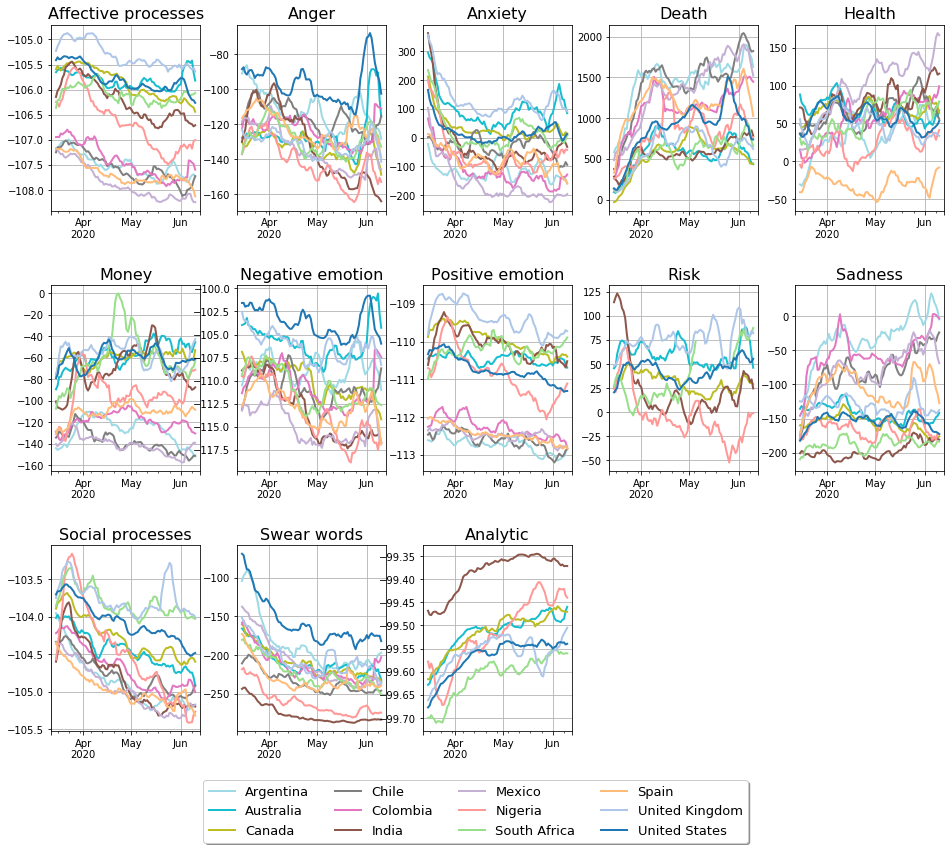}
    \caption{Time series for the NLSs for the countries as indicated by the legend. Each panel shows the individual linguistic categories. The units on the $y$-axis represent the percentage change of the National Linguistic Scores (NLS) on our data with respect to the LIWC baselines for Twitter (see Eq. (\ref{eq:liwc_score})). }
    \label{fig:liwc_panel}
\end{figure}

In Figure \ref{fig:liwc_panel}, we show the collection of NLSs for a selection of relevant linguistic categories. We observe clear trends that, in most cases, are synchronized between countries and languages. In particular, most categories associated with emotion -- notably Affect, Anger, Anxiety, Positive emotion, Negative emotion, and Swear words (swearing is associated with frustration and anger \cite{Jay2008}) -- have their highest scores in mid-to-late March, when the World Health Organisation (WHO) announced the pandemic status of Covid-19 and most Western countries introduced more stringent social restrictions \cite{SI}. These scores decay thereafter, indicating a relaxation of the emotional response in the conversation. This is consistent with results reported by Bhatia regarding the affective response to unexpected events \cite{bhatia2019affective} {  and with those of Aiello et al. \cite{aiello2020epidemic} where the Death NLS of the United States rises from late March on}. A qualitatively similar trend can be seen in the Social processes panel, the category involving ``all non-first-person-singular personal pronouns as well as verbs that suggest human interaction (talking, sharing)'' \cite{PennLIWC2015}. 

We also observe that health-related categories such as Death and Health show an overall rising trend, with Death rising most rapidly throughout March. These categories, with the exception of Positive Emotion and Health, peak again in the United States at the end of May, coinciding with the murder of George Floyd and the subsequent Black Lives Matter protests. Such universal trends are not apparent by visual inspection in the Money, Risk, and Sadness panels. An additional feature of these plots is the absolute scale of these values: in all cases, there is a significant percentage change from their baseline values, with large percentage increases observed initially in the use of words associated with Anxiety and later with Death, and a moderate percentage increase in the use of words associated with Risk.

\subsection{Comparing the public's perception with epidemiological data}

% New subsection in which we talk about System 1 and System 2, and about psychic numbing?

In this section, we explore the relationship between the NLSs described in Section \ref{PercLing}, which we use as a proxy for the public's perception, and the intensity of the pandemic, which we assume is the stimulus triggering this perception. Our measure of the intensity of the pandemic is the number of Covid-19 cases and deaths from the data described in Section \ref{EpData}.

\begin{figure}%[h!]
    \centering
    \includegraphics[width=1\linewidth]{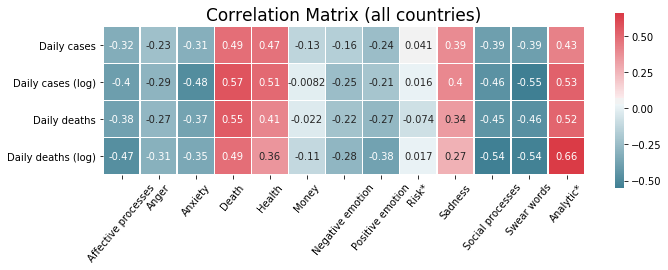}
    \caption{Correlation coefficients between epidemiological indicators and national linguistic scores (NLSs) averaged across all countries. *``Risk'' and ``Analytic'' are only available for the English-language LIWC. These two categories are thus averages across English-language countries only.}
    \label{fig:correlations}
\end{figure}

A straightforward way of approaching this relationship is by computing the correlations between the NLSs and the epidemiological data in a per-country basis, and we show the average across countries of these per-country correlations in Figure \ref{fig:correlations}. On the one hand, we observe significant negative correlations in emotionally charged categories (eg. Swear words, Anger, Anxiety, Affective processes), indicating a decay in emotion as the pandemic intensifies. Conversely, categories related with health and mortality (Death, Health) and analytical thinking (Analytic) show significant positive correlation\footnote{When analysing these correlations, we found that, overall, the cumulative cases and deaths correlate better with most linguistic categories than the daily data. However, while this is sensible in the early stages of the pandemic, it is unlikely to remain the case over a long time horizon due to humans' finite memory. We therefore proceeded with our comparison using the daily epidemiological data alone for this reason.}.

%FIGURE

%FIGURE

\subsubsection{Psychophysical numbing}\label{PsychNumb}

We believe the trends we observe in Fig. \ref{fig:liwc_panel} and the correlations we observe in Fig. \ref{fig:correlations} are consistent with the notion of {\bf psychophysical numbing}. This term was introduced by Robert Jay Lifton \cite{Lifton1982}, and developed by Paul Slovic \cite{Slovic2010, Fether1997} in the context of human perception of genocides and their associated death tolls, to describe the paradoxical phenomenon in which people exhibit growing indifference towards human suffering as the number of humans suffering increases. By inspecting the correlations between the NLSs and the epidemiological indicators, we find that as the pandemic intensifies -- in the sense of an increasing number of cases and deaths reported daily -- our emotional response diminishes, as expected from a psychophysical numbing phenomenon.

Specifically, we observe negative correlations between almost all components of the NLSs associated with affect -- Affective processes, Anger, Anxiety, Negative emotion, Positive emotion, and Swear words -- and the epidemiological data\footnote{The only exception is the cross-country average of the Sadness component of the NLSs, which is positively correlated with the epidemiological indicators and appears to be driven only from Argentina's, Chile's, and Colombia's increasing use of words related to Sadness. The remaining countries remain stationary at a lower-than-baseline value for this component.}. By inspecting Figure \ref{fig:liwc_panel}, we see that every country exhibits similar downward trends in these components and, with the exception of Anxiety, are all significantly lower than their baseline values throughout the observation period. 

This unusually low and decreasing Affect word count is accompanied, conversely, with a growing awareness of the morbidity of the situation in that we observe significant positive correlations between the Death NLSs and the daily national cases and deaths, indicating that the decrease in affect occurs simultaneously with and {\it despite} an attentional shift towards Covid-19 related mortality. We also observe a simultaneous increase in the Analytic component of each English-language dataset\footnote{Unfortunately, the Spanish LIWC dictionary does not yet have an Analytic category.} over this same period, indicating a movement towards more logical and analytical, rather than intuitive and emotional, thinking. 

The potential implication of this is that the public is less perceptive of the risk that the pandemic poses to public health, since their emotional response is reduced and reducing \cite{Sandman1993}. For example, Van Bavel {\it et al.} \cite{VanBavel} and Loewenstein {\it et al}. \cite{Loe2001} describe that risk perception is driven more by association and affect-based processes than analytic and reason-based processes, with the affect-based processes typically prevailing when there is disagreement between the two modes of thinking. The negative correlations between the intensity of the pandemic and affective processes, together with its positive correlation with the prevalence of analytic processes, suggests that public risk communication could be adjusted to re-balance the degree of affective and analytic thinking amongst members of the public to achieve favourable risk avoidance behaviour and, consequently, favourable public health outcomes.

\subsection{Analysing the emotional framing of Covid-19 casualties with semantic networks}%Word co-occurrence network analysis}
\label{sec:nw_analysis}

To support our claim that these observations are attributable to psychophysical numbing, we construct word co-occurrence networks using tweets in our dataset. {  Word co-occurrence networks are a class of  linguistic networks, in which nodes are words appearing in a body of text and an edge is placed between a pair of words with a weight given by some function of the number of co-occurrences of that pair in the text. Empirical word co-occurrence networks have been used in cognitive network science as approximate reconstructions of the author's latent cognitive structures, e.g. semantic or conceptual networks \cite{Siew2019}, with a given corpus deemed to be an empirical manifestation of such structures. 

For example, Kenett {\it et al.} \cite{Kenett2014} reconstruct participants' internal semantic networks on the basis of their responses in a free word association task, reporting that participants that were found independently to have lower creativity scores also had less well-connected semantic networks -- specifically, a higher modularity, average shortest path length, and diameter, and a lower small-world-ness \cite{Humphries2008} -- than participants scoring more highly in creativity. 
 
In the context of the Covid-19 pandemic, Stella {\it et al.} \cite{Stella} use word and hashtag co-occurrence networks in conjunction with word-to-emotion mappings to uncover complex emotional profiles amongst Twitter users posting from Italy during the first week of lockdown. More generally, a plethora of models for inferring semantic relationships between words in natural language processing tasks are based on some notion of word co-occurrence \cite{Jones2020}. The semantic proximity of a pair of words in such models has also been shown to possess predictive power regarding the subjective probability participants assign to hypothetical real-world events involving that pair of concepts \cite{bhatia2016vector}. For a more complete review of the use of linguistic networks in the study of human cognition, we refer the interested reader to \cite{Siew2019}.

Given the well-established utility of word co-occurrence analysis in providing a view of authors' internal cognitive structures, we employ such an approach on $\mathcal{W} = \text{Death } \cup \text{ Affect}$ -- the set of words in either the Death or Affect categories -- in an attempt to approximate the Twitter users' internal semantic relationships between these two concepts\footnote{{ The Affect category contains all the words related with affective processes. This includes the words in Anger, Anxiety, Positive and Negative emotion, and Swear words, which are all significantly correlated with Death and daily deaths.}}. Specifically, we hypothesise that, if the psychophysical numbing effect is legitimate, the modular structure of these networks will separate Death-related and Affect-related words more decidedly at larger daily death counts than at lower death counts. This would indicate that conversation regarding Covid-19-related mortality evokes a weaker emotional response at higher daily death counts.}

Given a set $\mathcal{T}$ of tweets, the word co-occurrence network $G(\mathcal{T})$ is represented by a weighted adjancency matrix $A(\mathcal{T})$ in which the nodes are words belonging to the Death and Affect LIWC dictionaries. Entry $A_{ij}(\mathcal{T})$ counts the number of co-occurrences between words $i$ and $j$ across all tweets in $\mathcal{T}$, and is computed as 
\begin{equation} 
    A_{ij}(\mathcal{T}) = \left(B(\mathcal{T})^T B(\mathcal{T}) \right)_{ij},
\end{equation}
where $B_{tk}(\mathcal{T})$ counts the number of instances of word $k$ in tweet $t \in \mathcal{T}$. We ignore self-edges by imposing $A_{ii} = 0$, since it is the relationship between distinct words that is of interest. (See Appendix \ref{DetailsCo} for further details on the construction of these networks.)

{  \subsubsection{Qualitative overview of the Death-Affect partition} }

\begin{figure}%[h!]
\centering
	\begin{subfigure}[b]{0.49\linewidth}
	\includegraphics[width=\textwidth]{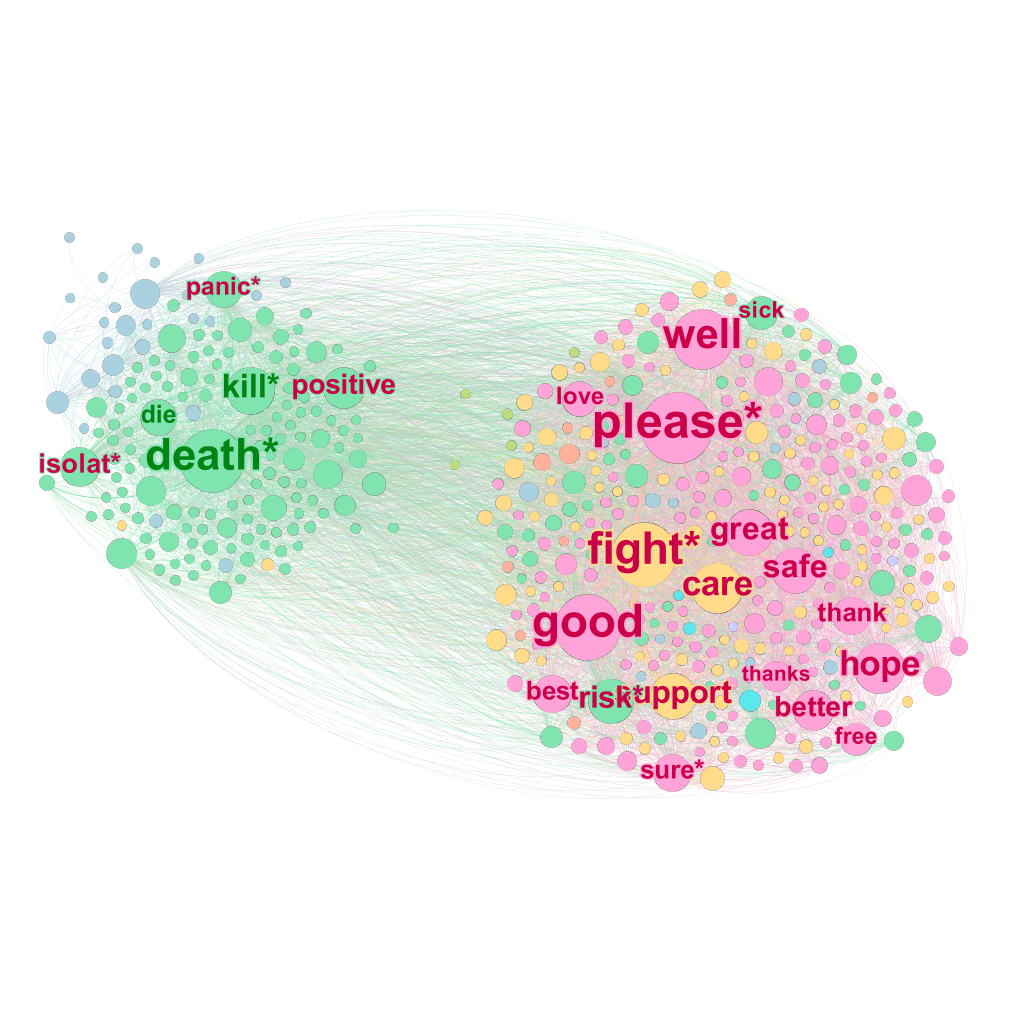}
	\caption{March 11th to April 9th, 2020. }	\label{fig:EngSnap1}
	\end{subfigure}
	\begin{subfigure}[b]{0.49\linewidth}	
	\includegraphics[width=\textwidth]{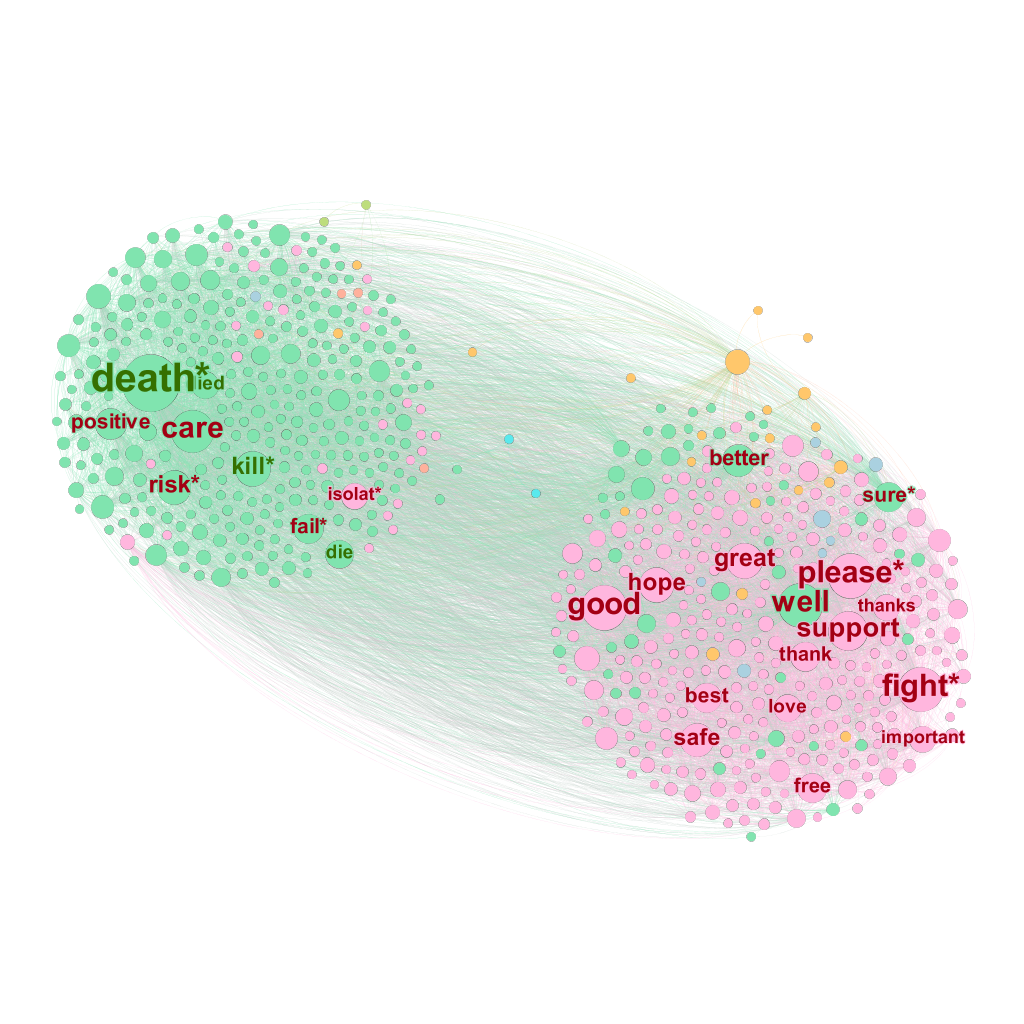}
	\caption{April 10th to May 23rd, 2020.}	\label{fig:EngSnap2}
	\end{subfigure}
	\begin{subfigure}[b]{0.49\linewidth}	
	\includegraphics[width=\textwidth]{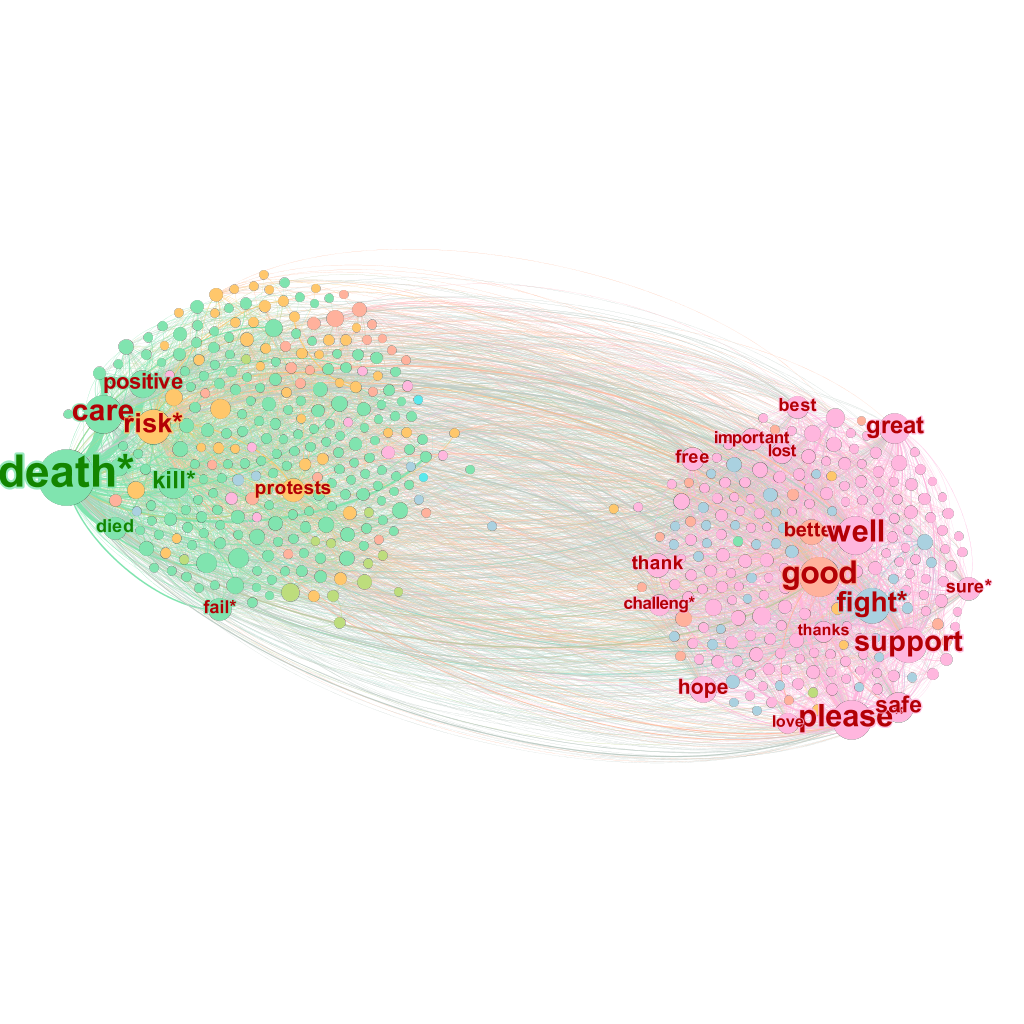}
	\caption{May 24th to June 13th, 2020.}	\label{fig:EngSnap3}
	\end{subfigure}
	\caption{Snapshots of the word co-occurrences associated with Death (green labels) and Affect (red labels) for English-language tweets aggregated across all analyzed countries in three different time windows (see sub-captions). The nodes are coloured according to their community label as obtained by maximising modularity with the Louvain algorithm \cite{Blondel_2008}. We filtered edges with weight below $20$ co-occurrences for visualisation purposes.}
\end{figure}

{  We identify three main periods for which we construct network snapshots of word co-occurrences (see \Cref{fig:EngSnap1,fig:EngSnap2,fig:EngSnap3}). The first period spans \textbf{11th March to 9th April 2020}, in which the WHO declared Covid-19's pandemic status and governments generally imposed social restrictions. The second period spans \textbf{10th April to 23rd May}, during which most Covid-19 cases either underwent exponential growth or flattened out for some countries in Europe. The final period spans \textbf{24th May to 13th June}, during which most countries were at the peak daily rate of Covid-19 cases or where in a stage of decreasing number of daily cases. Moreover, the Black Lives Matter protests were triggered by the murder of George Floyd in the USA in this period.} In constructing these networks, we weight each country equally by taking a random sample of approximately 300,000 tweets from each country.

{  In \Cref{fig:EngSnap1,fig:EngSnap2,fig:EngSnap3}, we visualise these three snapshots for the English-language tweets. From these we observe that two clusters emerge in all cases: a left-hand cluster consisting mainly of Death-related words and a right-hand cluster consisting primarily of Affect-related words. We also observe that the relative sizes of these clusters vary over time: the Death-cluster grows in size as the pandemic progresses, and remains separated from the Affect-based cluster. This indicates that the evolving structure of these networks may be consistent with our hypothesis of psychophysical numbing: throughout, Covid-19 casualties appear not to evoke a strong emotional response. 

However, we find that a number of the most highly connected nodes in these Death clusters are Affect-related words: in the first network, the Affect-related words ``panic'', ``positive'', and ``isolat*'' appear; in the second, the words ``care'', and ``fail*'' also appear; and in the third, ``protests'' appears. While such words are normally associated with affective processes, we argue that some of these are more readily understood in terms of their association with Covid-19-specific topics that are less indicative of an affective experience in this context than they might be more generally. For example, ``positive'' is used very frequently in the context of the pandemic in relation to individuals ``testing positive'' for the virus. In Table \ref{tab:AffectWords}, we address five of these words, providing what we believe are the most plausible explanations for their association with conversation surrounding mortality during the Covid-19 pandemic.

\begin{table}[]
\begin{center}
\begin{tabular}{lcl}
{\bf Word} & {\bf Figures} & {\bf Interpretation} \\ \specialrule{.1em}{.05em}{.05em}
positive & 3a to 3c & \makecell{Used in reference to the number of people that have tested positive\\ for Covid-19} \\
isolat*  & 3a, 3b & \makecell{Used in discussion surrounding symptomatic and at-risk individuals\\ self-isolating} \\
care     & 3b,3c & \makecell{Used in relation to: the health care system; the death care industry;\\ the admission of Covid-19 patients to intensive care units; and\\ deaths occurring in care homes for the elderly} \\
panic    & 3a & Panic-buying of household goods, e.g. toilet paper, hand-sanitiser  \\
protests & 3c & George Floyd's death and subsequent Black Lives Matter protests               
\end{tabular}\caption{Words belonging to the Affect LIWC category that appear in the primarily Death-based clusters in the three snapshot word co-occurrence networks shown in \Cref{fig:EngSnap1,fig:EngSnap2,fig:EngSnap3}. The middle and right columns indicate in which snapshots they are most prominent, and the likely explanation for their association with the concept of death during the pandemic.}\label{tab:AffectWords}
\end{center}
\end{table}
}

{  Altogether, this initial examination indicates that words associated with a subjective emotional/affective experience and words related to death may be well-separated in this Twitter data, which is consistent with the notion of psychophysical numbing as an explanation for the trends and correlations observed in Figures \ref{fig:liwc_panel} and \ref{fig:correlations}. For completeness, we include the equivalent co-occurrence graphs for the Spanish-language tweets in Appendix \ref{SpCo}, about which similar statements can be made. 

Our discussion has so far been qualitative given that the aforementioned network snapshots (i) vary considerably in size, (ii) represent the aggregate conversation of the tweets across countries in our dataset, and (iii) involve crude aggregation over large time periods. In the next section, we address these issues by investigating the change in a number of network measures over time and discuss the extent to which they support our hypothesis of psychophysical numbing.}\\

%FIGURE!

{  

\subsubsection{Quantitative analysis of the Death-Affect partition}

To further probe this hypothesis, we seek network measures that describe the strength of association between the concept of death and affective processes. Since the primary tenet of psychophysical numbing is that ``the more who die, the less we care'', our investigation is focused on the degree to which conversation around Covid-19 mortality evokes the use of affective language, which is our proxy for ``degree of caring''. In particular, we are interested in whether the emotional framing of such conversation changes as the daily death rates change in each country, where a less emotional conversation at higher daily death rates would support the hypothesis of psychic numbing.

For this purpose, we investigate the dynamics of the following network measures over a sequence of comparable snapshots for each country:
\begin{enumerate}
    \item the \textbf{weighted modularity} for the partition $\mathcal{P}_{\rm LIWC}$ induced by assigning nodes to their respective LIWC categories, i.e. Death or Affect. We define the weighted modularity following Newman \cite{Newman2004} as 
    \begin{equation}\label{eq:modularity}
        Q_{\rm LIWC}(t) = \frac{1}{2m(t)} \sum_{ij} \left( A_{ij}(t) - \frac{k_i(t) k_j(t)}{2m(t)} \right) \delta(c_i, c_j),
    \end{equation}
    where $A_{ij}(t)$ is the weighted adjacency matrix of a network at snapshot $t$, $k_i(t) = \sum_j A_{ij}(t)$ is the strength of node/word $i$, $m(t) = \frac{1}{2} \sum_{ij} A_{ij}(t)$ is the total strength of the network, $c_i \in \lbrace{\text{Death},\, \text{Affect}\rbrace}$ represents the community assigned of node $i$ under partition $\mathcal{P}_{\rm LIWC}$, and $\delta(\cdot, \cdot)$ is the Dirac delta function.
    
    \item the \textbf{fraction of the total strength of node ``{death*}'' ({``muert*''})} that can be attributed to its connections with other nodes in the Death category:
    \begin{equation}\label{eq:frac_Death}
        f_{\rm Death}(t) = \frac{1}{k_{{\rm death}^*}(t)}\sum_{j} A_{{\rm{death}^*}, j}(t)\, \delta(c_{{\rm death}^*}, c_j).
    \end{equation}
    The range of $f_{\rm Death}$ is bounded between $0$ and $1$, being $0$ when all of the neighbours of node ``death*'' (``muert*'') are in the Affect category and $1$ when all of its neighbours are in the Death category. 
\end{enumerate}

We henceforth omit the explicit time-dependence of $f_{\rm Death}$ and $Q_{\rm LIWC}$ to simplify notation. The first measure tracks the quality of separation of the Death and Affect categories in these semantic networks. A larger $Q_{\rm LIWC}$ indicates a better separation between Death and Affect in the empirical word co-occurrences. This is relevant to our investigation of psychophysical numbing in the following sense: if the numbing effect is genuine, we should expect that $Q_{\rm LIWC}$ is larger at larger values of the daily number of deaths and lower at lower values of the daily number of deaths. This would indicate that conversation around Covid-19-related deaths evokes affective responses less strongly for larger death rates. If a weakening association between the concept of death and affective processes is an accurate measure of growing apathy and indifference -- of the ``collapse of compassion'' \cite{Cameron2011} -- then observing a positive correlation between $Q_{\rm LIWC}$ and the daily national number of deaths would provide evidence supporting our hypothesis of psychophysical numbing.

The second is a local measure of the strength of association between the concept of Covid-19 deaths -- represented with the word ``death*'' (``muert*'') in a tweet -- and the affective processes within those tweets. A high $f_{\rm Death}$ value suggests a weak evocation of affective responses during conversation around Covid-19-related deaths.

To perform this analysis, we compute a sequence $\left(G_{t}\right)_{t}$ of higher-frequency snapshots than those in \Cref{fig:EngSnap1,fig:EngSnap2,fig:EngSnap3}, where $t = 1, \dots, T$ labels each of the $T$ snapshots for a given country. Each snapshot represents, on average, the tweets contained in 3 consecutive days and, for each country, each snapshot has roughly the same number of tweets (see Appendix \ref{DetailsCo} for details on the construction of these networks). With this construction, each network contains approximately the same number of nodes, edges, and network total strength, enabling a fair comparison of the network measures, Eqs. \ref{eq:modularity} and \ref{eq:frac_Death}, over time. 

In Figure \ref{fig:netMeas_panel}, we plot the $z$-scores of these network measures and of the log of the daily number of deaths $\log{ s(t)}$ for each country, and report the Pearson correlation coefficients $\rho_{\rm Q}$ and $\rho_{f}$ of $Q_{\rm LIWC}$ and $f_{\rm Death}$ with $\log{s(t)}$, respectively, in parenthesis above each plot. In general, we observe similar dynamics for both $Q_{\rm LIWC}$ and $f_{\rm Death}$. This is sensible, since both are measures of the relative strength of association within the two communities induced by the Death and Affect word sets. Furthermore, we observe a number of instances -- most notably, Canada, Colombia, Mexico, the United Kingdom, and the United States -- in which there is a relatively strong correlation between these network measures and $\log{s(t)}$. These correlations are, however, weaker in other countries to varying degrees.

%FIGURE!
\begin{figure}%[h!]
    \centering
    \includegraphics[width=1\linewidth]{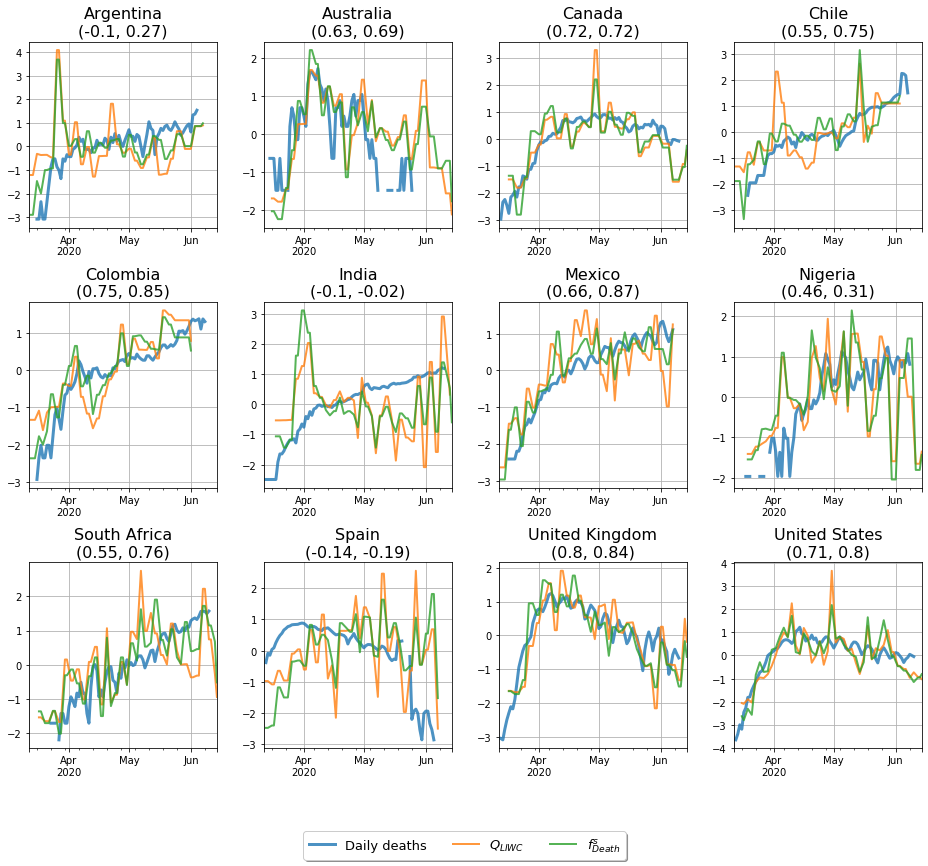}
    \caption{{ Panel plot time series for the network measures $Q_{\rm LIWC}$ and $f_{\rm Death}$ (see Eqs. (\ref{eq:modularity}) and (\ref{eq:frac_Death}) respectively).}}
    \label{fig:netMeas_panel}
\end{figure}

To verify that the observed $\rho_{\rm Q}$ and $\rho_{f}$ can be attributed to the empirical word co-occurrences, 
we compute the same correlations for corresponding sequences of null network models. For our null model, we take the weighted version of the configuration model described in \cite{Britton2011}. Here, a realisation $G_{j,t}^{\rm null}$ of the null model at random seed $j$ involves assigning node $i$ $D_i$ stubs, where
\begin{equation}
    D_i \sim p(d)
\end{equation}
and $p(d)$ is the empirical degree distribution. The $k$th stub for node $i$ is then assigned a weight
\begin{equation}
    W_{ik} \sim p(w | D_i = d ),
\end{equation}
where $p(w | D_i = d)$ is the empirical distribution of weights $W$ for nodes with degree $w$. Stubs with the same weight are then joined with uniform probability. 

As a baseline, we compute a sequence $\left( \left[G_{j,t}^{\rm null} \right]_j \right)_{t}$
%\begin{equation*}
%    \left( \left[G_{j,t}^{\rm null} \right]_j \right)_{t}
%\end{equation*}
of null model ensembles for each country at each snapshot $t$, where $j = 1, \dots, J$ labels each of the $J$ realisations of the null model. Here, we take $J = 100$ realisations per snapshot. We then compute the average network measure over each ensemble for both $Q_{\rm LIWC}$ and $f_{\rm Death}$, here denoted $Q_{\rm LIWC}^{\rm null}$ and $f_{\rm Death}^{\rm null}$ respectively. Similarly, we write the correlation coefficients as $\rho^{\rm null}_{\rm Q}$ and $\rho^{\rm null}_{f}$. We report these coefficients, along with the correlation coefficients for the empirical networks, in Figure \ref{fig:netMeas_correlations}.

%FIGURE!
\begin{figure}%[h!]
    \centering
    \includegraphics[width=1\linewidth]{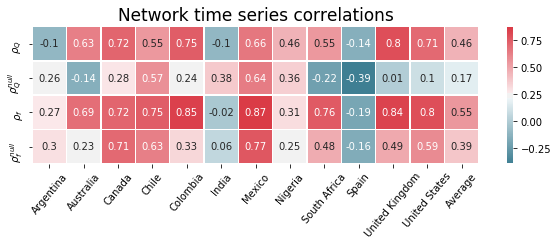}
    \caption{ { Correlation coefficients between the daily national death counts and the network measures $Q_{\rm LIWC}$ and $f_{\rm Death}$ (see Eqs. (\ref{eq:modularity}) and (\ref{eq:frac_Death}) respectively). } }
    \label{fig:netMeas_correlations}
\end{figure}

We find that, in most cases, the correlation coefficients are higher for the empirical word co-ocurrences than for the null model counterparts. In particular, each of Australia, Canada, Colombia, South Africa, the United Kingdom, and the United States have $\rho_{\rm Q} \gg \rho^{\rm null}_{\rm Q}$. This difference is also present for Nigeria, although it is smaller in this instance. For the remaining countries, however, the differences are either negligible or in the opposite direction. Overall, nonetheless, we see that the average across countries of ${\rho}^{\rm null}_{\rm Q}$ is low, whereas the average across countries of ${\rho}_{\rm Q}$ is almost three times larger, and that $\rho_{\rm Q} > \rho^{\rm null}_{\rm Q}$ for nine cases out of twelve.

A similar pattern is observed for $\rho_{f}$ and $\rho^{\rm null}_{f}$. In some instances -- namely Australia, Colombia, South Africa, the United Kingdom, and the United States -- we observe an increase $> 0.2$ in the correlation coefficients of the original sequence of snapshots relative to the corresponding sequences of null snapshots. This indicates that these increases in $\rho_{f}$ can be attributed to the empirical word co-occurrences. 
The difference is smaller but nonetheless in the correct direction for Chile, Mexico, and Nigeria. For the remaining countries, the difference is negligible.

\subsubsection{Discussion and summary}

Overall, this analysis provides evidence in favour of our hypothesis of psychophysical numbing, although this evidence is not definitive. We have seen that, for most countries, the separation between words associated with Death and Affect in our approximate semantic networks -- as measured by $Q_{\rm LIWC}$ and $f_{\rm Death}$ -- becomes more pronounced as the national daily deaths rise, and that this relationship is generally weaker in the null model realisations. 

There are nonetheless some exceptions to this statement. In particular, we find for Chile and Mexico that the difference between $\rho$ and $\rho^{\rm null}$ is marginal, but that both versions of the correlation coefficients are high. We also report low correlations between these network measures and the time series of daily deaths for Argentina, India, and Spain. For the case of Spain, however, there are two exogenous death-related events contributing to this anomalous behaviour and low correlation values, see Appendix \ref{app:peaks} for details. For the case of India, there is evidence suggesting that Twitter users posting from India have a strong preference for using Hindi in the expression of negative sentiment and emotion, but English in the expression of positive emotion \cite{rudra2016understanding}. Our use of an English-language dictionary for evaluating the emotional content of such tweets may therefore bias our results, and a more thorough analysis including tweets and dictionaries in both Hindi and English (or in ``Hinglish'', the blending of the two \cite{Mathur2019}) should be performed in future. This is a specific case of a more general problem regarding the use of a single dictionary to analyse texts from different world regions, which typically differ in dialect.

For the remaining countries in our dataset, however, the empirical co-occurrences yield stronger correlations between the network measures and the national daily deaths than in the case of the baseline models, providing support for our psychophysical numbing hypothesis. Our observations thus indicate that psychophysical numbing may be a genuine effect for many Twitter users, but that other factors are possibly contributing to our results. 
% about 20% Nigerians on Twitter: https://gs.statcounter.com/social-media-stats/all/nigeria
% 20-30% UK, from same website
% 10-20% US, same website
Some of these factors are methodological issues with this work. First, we saw in \Cref{fig:EngSnap1,fig:EngSnap2,fig:EngSnap3} that LIWC is unable to account for context, and that there are a number of words that are classically associated with affective processes that are more appropriately associated with concepts surrounding mortality in the context of the pandemic. Second, in analysing word co-occurrences, we only retain tweets that contain at least two distinct words in the set Death $\cup$ Affect by construction. We have evaluated separately the proportion of tweets in each snapshot that contribute to our word co-occurrence networks, and have seen that this usually corresponds to between 10-20\% of tweets for each snapshot, with between 20-30\% of tweets involving the use of only one word in Death $\cup$ Affect. As such, this potentially leads to a systematic overestimation of the relative strength of association between words in Death $\cup$ Affect. Finally, as with most studies of organic social media data, it is hard to control for exogenous factors that form part of the Covid-19 conversation (e.g. Black Lives Matter protests, death-related news). It is thus important to treat such evidence as complementary to classical laboratory-based, controlled psychological experiments.

}

\subsection{Modeling attention to Covid-19 casualties}

In the previous section, we demonstrated our finding that as the pandemic intensifies, the proportion of words that appear in the set of Tweets posted in each country that indicate emotion diminishes over time. This indicates that the actual emotional response to the pandemic diminishes as the intensity of the pandemic increases, implying a psychophysical numbing effect. We supported this explanation by showing that the word co-occurrence networks induced by our set of tweets host a community structure that separates words in the Death and Affect dictionaries, suggesting that people do not talk about Covid-19 deaths in a highly emotional tone. {  We built on this analysis by tracking a number of measures of this supposed separation in higher-frequency sequences of snapshots for each country, observing that these network measures behaved consistently with our hypothesis of psychophysical numbing for a number of countries.}

The following sections model the relationship between the progression of the Covid-19 pandemic and the Twitter users' perception using grounded theories of psychophysical numbing. {  Until this point, we have used the emotional framing of the conversation around Covid-19 mortality as an indication of the degree of concern or indifference towards these casualties. However, one could argue that attention itself is equally indicative of the degree of concern experienced by individuals regarding such casualties. Indeed, both are recognised as key components to risk perception and the perception of threats \cite{Slovic2010}. For this reason, we investigate the relationship between the typical perceptual response of individuals to a stimulus, in this case the daily number of reported deaths nationally, and seek to describe this relationship using established psychophysical laws, as in previous lab-based psychological experiments e.g. \cite{Stevens1957}.} 

\subsubsection{The Weber-Fechner law}\label{WFL}

\begin{figure}%[h!]
    \centering
    \includegraphics[width=1\linewidth]{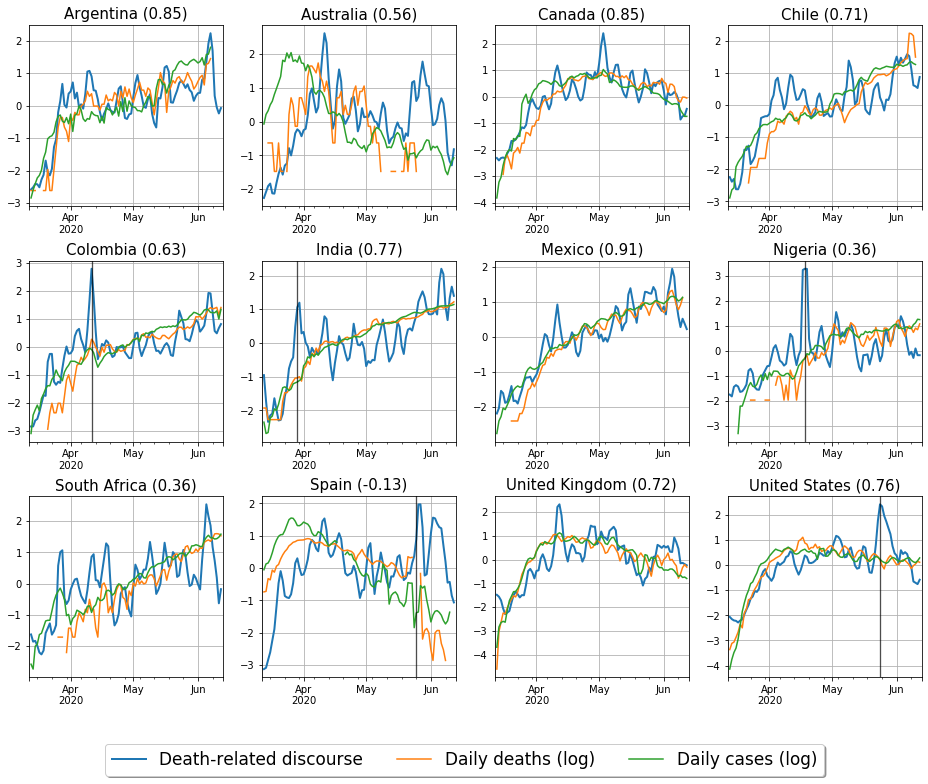}
    \caption{
    Panel time series for $p_i^{\text{Death}}(t)$ (blue), the logarithm of the daily deaths (orange), and the logarithm of the daily cases (green). Each panel presents a different country, with the country name provided in the subplot title. The correlation between $p_i^{\text{Death}}(t)$ and the national daily death rate is given in parentheses for each country. Data is smoothed with a 3-day moving average and standardized with their $z$-score to make them visually comparable. Vertical lines represent peaks in the death discourse caused by exogenous events {  not related to psychophysical numbing (see Appendix \ref{app:peaks} for details)} which we remove from the time series.  
    }
    \label{fig:liwc_vs_data}
\end{figure}

Our analysis suggests that the public's perception of the progression of the pandemic is logarithmic or, at least, sublinear. From Figure \ref{fig:correlations}, we observe that the correlation magnitudes between NLSs and epidemiological data are generally larger in absolute value whenever the latter are taken in logarithmic scale. To exemplify this observation, we show in Figure \ref{fig:liwc_vs_data} the $z$-scores\footnote{Recall that the $z$-score of a sequence of observations $\mathbf{Y} = (y_1, \cdots, y_T)$ is given by $\mathbf{Z} = (\mathbf{Y} - \mu_Y)/\sigma_Y$, where $\mu_Y$ and $\sigma_Y$ are the mean and standard deviation of $\mathbf{Y}$, respectively.} of the Death NLSs and of the logarithm of the daily number of deaths and cases within each country. 
 
 %FIGURE

The general correspondence between all three normalised features in each country is striking\footnote{We note that the correspondence is weaker for Australia, Nigeria, and South Africa due to the relatively low number of cases in these countries (see Fig. \ref{fig:covid_data} in the Appendix for reference). The correspondence is also weaker in Spain {  because it contains two exogenous peaks not related to psychophysical numbing. See Appendix \ref{app:peaks} for a discussion of these peaks for Spain and other countries, which we remove from the time series}.}
% , for two reasons: due to its revision of the number of cases in late May, resulting in a day of ``negative deaths''; and due to their having recorded a day with no Covid-19-related deaths, which was a significant event given that Spain had seen many deaths until that point.}. 
We propose that this can be explained in terms of the {\bf Weber-Fechner law} \cite{fechner1966elements}, which is a quantitative statement with its origins in psychology and psychophysics regarding humans' perceived magnitude $p$ of a stimulus with physical magnitude $s$. It states that a human's perception of the magnitude of a stimulus varies as the logarithm of the physical magnitude $s$ of the stimulus, meaning we are more sensitive to ratios when comparing different physical magnitudes than we are to absolute differences. In the continuum limit, Eq. (\ref{eq:WFL_}) gives the following functional form for the Weber-Fechner law:
\begin{equation}\label{eq:WFL}
    p(t) = k \log{ \frac{s(t)}{s_0} } + R(t),
\end{equation}
where $k$ and $s_0$ are real-valued parameters and $R(t)$ the residual. Parameter $k$ determines the sensitivity of perception to changes in the stimulus $s$, while $s_0$ determines the minimum threshold that the stimuli $s$ must overcome in order to be perceived. The residual term $R(t)$ is a random variable representing noise not directly captured by the stimulus. For instance, exogenous events can trigger abrupt peaks in the Death score. This is the case, for example, with the murder of George Floyd in the United States, or the peak in Nigeria around April 17th 2020, triggered by a number of prominent African figures dying from Covid-19 around that day, including the Nigerian President's top aide (see Appendix \ref{app:peaks} for details in these peaks).

In order to test the Weber-Fechner law, we fit a linear regression model to $p^{\rm Death}_i(t)$, the Death NLS time series in country $i$, and $\log{s_i(t)}$, the daily number of deaths in the same country, and summarize the results of these fits in Table \ref{tab:WFL-fit}. We find that Eq. (\ref{eq:WFL}) accurately models the data, with significant coefficients ($p$-value $< 0.01$) for all countries except Spain. %? EVEN WITH END DATA REMOVED?)
The sensitivity parameter $k$ has the same order of magnitude for all significant countries. However, the country with the lowest $k$ is $\sim 3$ times less sensitive than the highest, indicating that Twitter users in different countries may react differently to the evolution of the pandemic. The minimum stimuli threshold $s_0$, in the other hand, is always small: most countries, except for the United States and the United Kingdom, need only one Covid-19 death in a given day in order to be perceived. Conversely, the United States and United Kingdom need approximately $5$ and $6$ deaths to be perceived, which is small compared to the thousands of daily deaths registered in these countries during the observation period.

\begin{table}%[h!]
\centering
\resizebox{\textwidth}{!}{\begin{tabular}{l|cc||ccc||cccc}
\toprule
{Weber-Fechner law}               & $k$ & $s_0$ & 95\% CI &  $t$  &  $P>|t|$ & $R^2$ &    NRMSE & $n$ \\ %  AIC & 
\textbf{Country} &     &       & ($k$)   & ($k$) &   ($k$)  &       &          &     \\ %  AIC & 
\midrule
\textbf{Argentina     } &  1.044 &   0.0080 &   0.758 -- 1.329 &   7.29 &     0.0 &  0.421 &  0.113 &  75 \\ %  44.9 & 
\textbf{Australia     } &  1.042 &   0.1047 &   0.508 -- 1.576 &   3.94 &  0.0003 &  0.275 &  0.171 &  43 \\ %  29.8 & 
\textbf{Canada        } &  0.596 &   0.4477 &   0.508 -- 0.683 &  13.53 &     0.0 &  0.683 &  0.105 &  87 \\ %  -11.9 & 
\textbf{Chile         } &  0.575 &   0.0001 &   0.412 -- 0.737 &   7.05 &     0.0 &  0.395 &  0.149 &  78 \\ %   83.6 & 
\textbf{Colombia      } &  0.604 &   0.0011 &   0.436 -- 0.772 &   7.18 &     0.0 &  0.414 &  0.144 &  75 \\ %   32.6 & 
\textbf{India         } &  0.332 &   0.0061 &     0.264 -- 0.4 &   9.69 &     0.0 &  0.543 &  0.128 &  81 \\ %  -36.4 & 
\textbf{Mexico        } &  0.846 &   0.0300 &    0.742 -- 0.95 &   16.2 &     0.0 &  0.775 &  0.101 &  78 \\ %   38.8 & 
\textbf{Nigeria       } &  0.457 &   0.0003 &   0.168 -- 0.747 &   3.16 &  0.0025 &  0.147 &  0.222 &  60 \\ %   38.8 & 
\textbf{South Africa  } &  0.282 &   0.0001 &   0.127 -- 0.436 &   3.64 &  0.0005 &  0.171 &  0.183 &  66 \\ %   33.5 & 
\textbf{Spain         } & -0.016 &        inf &  -0.198 -- 0.165 &  -0.18 &  0.8593 &    0 &  0.198 &  82 \\ %  152.3 & 
\textbf{United Kingdom} &  0.752 &   5.241 &    0.61 -- 0.894 &  10.54 &     0.0 &  0.555 &   0.143 &  91 \\ %  53 & 
\textbf{United States } &  0.788 &   4.2478 &   0.672 -- 0.905 &  13.43 &     0.0 &  0.677 &  0.126 &  88 \\ %  57.8 & 
\bottomrule
\end{tabular}}
\caption{Results from the fit of the Weber-Fechner law to the observed relationship between the Death NLS and the logarithm of the daily number of deaths in each country (see Figure \ref{fig:liwc_vs_data}). Overall, this model best describes the relationship between the daily number of deaths local to each country and the Death NLS.}
\label{tab:WFL-fit}
\end{table}

\subsubsection{Power-law perception}\label{PLP}

An alternative functional form for the relationship between human perception $p$ of a stimulus and the physical magnitude $s$ of the stimulus is a power law relationship
\begin{equation}
    p(t) = \nu\cdot s(t)^{\beta} + \Tilde{R}(t),
    \label{eq:powerlaw}
\end{equation}
where $\nu$ and $\beta$ are parameters determining the perception from a stimulus of unit magnitude and the growth rate of the perception as a function of the stimulus magnitude, and $\Tilde{R}(t)$ is a residual term. This form has been shown to outperform the Weber-Fechner law in characterising human perception in a number of empirical studies \cite{Stevens1975}. We also therefore report the results of this model fit to the relationship between the Death NLS $p^{\rm Death}_i(t)$ and national daily death counts $s_i(t)$ for each country $i$, reporting our results in Table \ref{tab:PL-fit}.

In all cases, we observe sublinear exponents $\beta$ for the perception of the daily deaths data, with significant exponents ($p$-value $< 0.01$) ranging between $0.085$ and $0.36$. These exponents are of the same order of magnitude as the $\beta$ of $0.32$ reported in \cite{Summers1994}, where in several laboratory experiments they measure psychophysical numbing in participants' perception of death statistics. As discussed previously, the data for Spain is unusual for a number of reasons, thus the model does not accurately describe the data in this instance. These results suggest that Twitter users in certain countries are more sensitive to change in the number of deaths than others.

%TABLE
\begin{table}%[h]
\centering
\resizebox{\textwidth}{!}{\begin{tabular}{l|cc||ccc||cccc}
\toprule
{Power law}               & $\beta$ & $\nu$ &   95\% CI &       $t$ &   $P>|t|$  & $R^2$*  &    NRMSE & $n$ \\ %  AIC & 
\textbf{Country} &         &       & ($\beta$) & ($\beta$) & ($\beta$)  &       &          &     \\ %         &
\midrule
\textbf{Argentina     } &  0.164 &       2.21 &  0.121 -- 0.208 &   7.59 &     0.0 &  0.411 &   0.114 &  75 \\ % 46.2 & 
\textbf{Australia     } &  0.363 &       0.99 &  0.181 -- 0.546 &   4.02 &  0.0002 &  0.259 &   0.173 &  43 \\ % 30.8 & 
\textbf{Canada        } &  0.288 &       0.37 &  0.252 -- 0.323 &  16.29 &     0.0 &  0.678 &   0.106 &  87 \\ % -10.4 & 
\textbf{Chile         } &  0.085 &       2.47 &   0.06 -- 0.109 &   6.97 &     0.0 &  0.382 &   0.151 &  78 \\ %  85.3 & 
\textbf{Colombia      } &  0.112 &       1.81 &  0.083 -- 0.142 &   7.57 &     0.0 &  0.425 &   0.143 &  75 \\ %  31.3 & 
\textbf{India         } &  0.126 &       0.77 &   0.101 -- 0.15 &  10.33 &     0.0 &  0.558 &   0.126 &  81 \\ %  -39 & 
\textbf{Mexico        } &  0.141 &       1.52 &  0.126 -- 0.157 &  18.04 &     0.0 &   0.78 &     0.1 &  78 \\ %  37.2 &  
\textbf{Nigeria       } &  0.104 &       1.56 &  0.037 -- 0.172 &   3.09 &  0.0031 &  0.143 &   0.223 &  60 \\ %  45.4 & 
\textbf{South Africa  } &  0.087 &       1.11 &  0.037 -- 0.136 &   3.52 &  0.0008 &   0.16 &   0.184 &  66 \\ %  34.4 & 
\textbf{Spain         } &  0.014 &       2.14 &  -0.03 -- 0.059 &   0.64 &  0.5241 & -0.042&   0.202 &  82 \\ %  155.7 & 
\textbf{United Kingdom} &  0.356 &       0.16 &  0.302 -- 0.409 &  13.21 &     0.0 &  0.514 &   0.149 &  91 \\ %  61.1 & 
\textbf{United States } &  0.309 &       0.21 &  0.279 -- 0.339 &  20.54 &     0.0 &  0.608 &   0.139 &  88 \\ %  74.8 & 
\bottomrule
\end{tabular}}
\caption{ The results from the fit of a power law to the relationship between the Death NLS and the national daily death count. This is the best model in some cases, though is outperformed by the Weber-Fechner law most times. 
*While we fit this model assuming a log-log relationship between $p$ and $s$, we compute $R^2$ with linear $p$ to make it comparable to the model implied by the Weber-Fechner law (see Eq. (\ref{eq:rsquared}) in Appendix \ref{app:comparison} for details). This may cause negative values of $R^2$ as is the case for Spain.}
\label{tab:PL-fit}
\end{table}

\subsubsection{Model comparison}

Both the Weber-Fechner law and power-law relationships between the Death NLS and the daily number of reported deaths accurately model the data. Each captures the phenomenon in which ``the first few fatalities in an ongoing event elicit more concern than those occurring later on'' \cite{Olivola2015}. By way of comparison, we present in Table \ref{tab:NRMSE}
the normalised root mean squared errors (NRMSE), defined as
\begin{equation}
    \text{NRMSE} = \frac{ \sqrt{ \frac{1}{n} \sum_t^n e(t)^2 } }{p_{max} - p_{min} },
    \label{eq:nrmse}
\end{equation}
for these models, in addition to a linear model between $p_i^{\rm Death}(t)$ and $s_i(t)$ as a baseline ``null'' model. Here, $ e(t) = p(t) - \hat{p}(t) $ is the model residual, and $n$ is the sample size. The models are directly comparable in this sense, since each involves only two parameters. Bhatia \cite{bhatia2016vector} performed a similar model comparison to test psychophysical laws for subjective probability judgements of real-world events, in that case finding that the linear relationship was the best. In our case, however, a linear relationship between $s$ and $p$ is significantly worse than the present concave models of perception (see Appendix \ref{app:comparison} for the results of the linear model), reinforcing our hypothesis of psychophysical numbing. 
%TABLE
\begin{table}%[h!]
\centering
% \resizebox{\textwidth}{!}
{\begin{tabular}{l|ccc}
\toprule
{NRMSE} & Power law & Weber-Fechner law & Linear relationship \\
\textbf{Country       } &           &             &                     \\
\midrule
\textbf{Argentina     } &     0.114 &       \textbf{0.113} &               0.116 \\
\textbf{Australia     } &     0.173 &       \textbf{0.171} &               0.175 \\
\textbf{Canada        } &     0.106 &       \textbf{0.105} &               0.117 \\
\textbf{Chile         } &     0.151 &       \textbf{0.149} &                0.17 \\
\textbf{Colombia      } &     \textbf{0.143} &       0.144 &               0.145 \\
\textbf{India         } &     0.126 &       0.128 &               \textbf{0.125} \\
\textbf{Mexico        } &       \textbf{0.1} &       0.101 &               0.133 \\
\textbf{Nigeria       } &     0.223 &       0.222 &               \textbf{0.218} \\
\textbf{South Africa  } &     0.184 &       \textbf{0.183} &               0.188 \\
\textbf{Spain         } &     0.202 &       0.198 &               \textbf{0.193} \\
\textbf{United Kingdom} &     0.149 &       \textbf{0.143} &               0.166 \\
\textbf{United States } &     0.139 &       \textbf{0.126} &               0.179 \\ \hline
\textbf{Mean          } &     0.151 &       \textbf{0.149} &                0.16 \\
\textbf{Proportion of best fits } &   16.7 \% &     \textbf{58.3 \%} &               25 \% \\
\textbf{Proportion of second-best fits } & \textbf{66.7} \% &   33.3 \% &               0 \% \\
\bottomrule
\end{tabular}}
\caption{ Comparison of the normalised root mean squared error (NRMSE) (see Eq. (\ref{eq:nrmse})) between the power law model of Eq. (\ref{eq:powerlaw}), the Weber-Fechner model of Eq. (\ref{eq:WFL}), and a linear relationship between variables, which we use as a benchmark model. Lower values indicate better-fitting models. Note that, overall, the Weber-Fechner law outperforms the other models. For further details, see Figs. \ref{fig:scatterplot_weber} and \ref{fig:scatterplot_powerlaw} in Appendix \ref{app:comparison}.}
\label{tab:NRMSE}
\end{table}

While the Weber-Fechner law is better than the power law model overall, the difference in their goodness of fit -- as measured by the NRMSE -- is marginal. Both are reasonable descriptions of the observed relationship, and similar conclusions can be drawn from both. 

\begin{figure}[h!]
    \centering
    \includegraphics[width=0.7\linewidth]{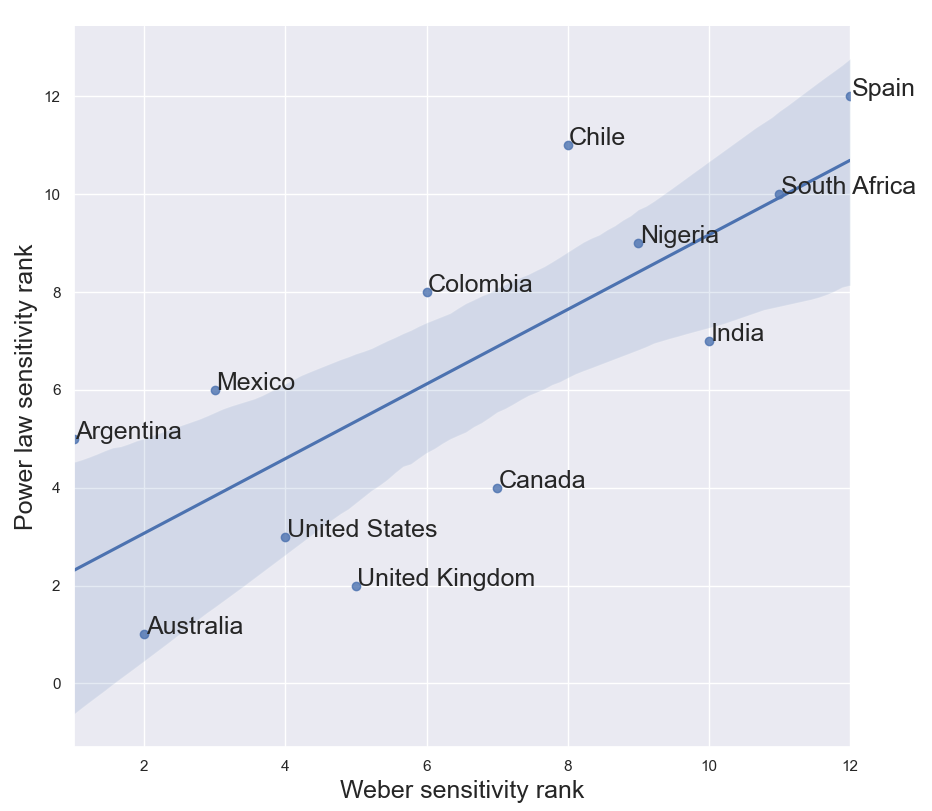}
    \caption{Comparison of the rank of each country as determined by their $k$ and $\beta$ parameters in the Weber-Fechner and power-law fits, respectively, which determine the sensitivity of Twitter users tweeting from each country to changes in the number of daily reported deaths. Low rank indicates high sensitivity relative to the remaining countries. The correlation between countries' ranks from both measures is high at 0.77.}
    \label{fig:rank_scatter}
\end{figure}

In particular, the parameters $k$ and $\beta$ from the Weber-Fechner law and power law, respectively, are analogous in their interpretation as the measure of the sensitivity of the nation's Twitter users to changes in the national Covid-19 daily death rate. To illustrate this, we rank the countries in our dataset in order of sensitivity to changes in the local death rate, as measured separately by these two parameters, and plot the correlation between the countries' ranks in Figure \ref{fig:rank_scatter}. Here, low rank indicates high sensitivity to changes in the number of daily deaths nationally. The correlation between the two methods of ranking -- according to $k$, the Weber-Fechner law slope parameters, and according to $\beta$, the power law model exponents -- is high, with correlation coefficient 0.77. This shows that the sensitivity of each country is relatively robust between models. By both measures, therefore, Twitter users tweeting in English and Spanish from Australia and Argentina, respectively, appear to be the most sensitive to changes in the national daily death rate, while Twitter users posting in English from South Africa, India, and Nigeria and in Spanish from Spain and Chile appear to be the least sensitive to these changes. 

\section{Discussion and conclusions}\label{Conc}

We explored the country-by-country relationship between the linguistic features present in a large set of tweets posted in relation to the Covid-19 pandemic, and the progression/intensity of the pandemic as measured by the daily number of cases and deaths in each country we consider. By considering the change, relative to a baseline, in the percentage of words present in each tweet that are associated with a number of psychologically meaningful categories -- here called linguistic scores -- we observed significant trends that we believe are indicative of a psychophysical numbing effect \cite{Slovic2010}. 

We found that the national linguistic scores (NLSs, see Eq. (\ref{eq:liwc_score})) associated with emotion and affect decrease as the pandemic intensifies. This is in spite of a greater attentional focus on death and mortality and a simultaneous increase in use of words indicating analytic reasoning. We showed, by constructing word co-occurrence networks on different time periods of the pandemic, that words related to death co-occur more frequently with other words related to death than they do with words indicating affect and emotion{.   We constructed network measures of this separation between the concepts of death and emotion -- namely the weighted modularity of the partition induced by the Death and the Affect LIWC dictionaries, and the fraction of strength of the ``death*'' (muert*) node attributable to connections with other nodes in the Death category -- and showed that this separation became more pronounced at larger daily death rates for a number of countries}. This is consistent with the notion of psychophysical numbing, which we believe may explain these observations. 

We also showed that the psychophysical laws of Weber-Fechner and of power law perception in humans accurately model the relationship between the frequency of words related to death and the actual daily number of Covid-19 deaths in each country. We estimated sub-linear exponents in the power law perception function that are of similar values to values previously estimated from psychological experiments \cite{Summers1994}. These exponents, together with parameter $k$ of the Weber-Fechner law (see Eq. (\ref{eq:WFL})), tell us how sensitive the Twitter users in each country are to their national Covid-19 daily deaths, and were seen to vary by country, indicating inter-country differences in risk perception and sensitivity to death rates. Such sensitivities were consistent across models (see Fig. \ref{fig:rank_scatter}) suggesting that these measures of a nation's Twitter users' sensitivities to changes in the national death rate are robust features of the data. 

{  Overall, our results indicate that two key factors contributing to risk perception -- attention and emotion \cite{Slovic2010} -- may be evolving in line with that predicted by psychophysical numbing amongst members of the public. In general, both measures of the degree of concern towards Covid-19-related casualties expressed by the Twitter users in our dataset appear to decrease as the number of Covid-19-related casualties increases. This potentially reflects a collapse of compassion and a concavity in the value assigned to human lives as the number of potential casualties grows.}

% Further, we show that the psychophysical laws of Weber-Fechner and of power law perception in humans accurately model the relationship between the frequency of words related to death and mortality and the actual daily number of Covid-19 deaths in each country. We estimate exponents in the power law perception function that are of similar values to values previously estimated from psychological experiments \cite{Summers1994}. These exponents, together with parameter $k$ of the Weber-Fechner law (see Eq. (\ref{eq:WFL})), are indicative of how sensitive the Twitter users in each country are to their national Covid-19 daily deaths. We find such sensitivities to be consistent across models (see Fig. \ref{fig:rank_scatter}), suggesting that the findings are robust features of the data.

Our findings illustrate the signaling power of Twitter, and demonstrate its potential use as a tool for monitoring public perception of risk during large-scale crisis scenarios. With the modelling and visualisation approaches we employ in this paper, policy-makers and public officials could track in near-to-real-time the public's attitudes towards threats to public well-being and the prevalence of factors important to public perception of risk, including degree of outrage and relative attentional focus on the threat. Our findings also imply a functional form for agent perception of the system state in models of opinion dynamics. This will be instrumental for developing coupled opinion dynamics-epidemiological models, in which the bidirectional relationships between human perception, human behaviour, and epidemic progression are modelled endogenously. 

A natural extension to this work would involve nowcasting and/or forecasting of certain economic indicators. It has also been limited in that we assumed that only the national death rate is a significant predictor of perception. A more complete analysis should account for the effect of other countries' death statistics as a driver of local perception, or more broadly an advancement of a process-level explanation of the cross-cultural differences we observe in the sensitivity to death statistics. This analysis could also be enhanced by relating these measures of risk perception to behavioural data, which -- since ``people's behavior is mediated by their perceptions of risk'' \cite{weber2004perception} -- may be useful for understanding the role of emotions in driving behaviours that are conducive to public health during crises. Further, a deconstruction of the aggregate indicators we have developed to the state and regional level may be necessary to more accurately characterise the relationship between local crisis progression and human risk perception. 

{  It is important to acknowledge that additional factors may be at play and contributing to our findings. In particular, our dataset is a large social media dataset in which non-human accounts -- for instance, bots, institutional accounts, and companies' public relations accounts -- coexist with human accounts. Such public relations and institutional accounts can be subject to editorial constraints on the kind of language used, and therefore may not reflect any true underlying subjective experience. The use of tweets from such non-human accounts may nonetheless be appropriate. Indeed, it is widely accepted that news media play a significant role in shaping public attention and opinion, e.g. via the Cultivation or Agenda-Setting theories of consumer-media relations \cite{Cultivation, AgendaSetting}. With almost half of all UK adults consuming news through social media in 2020 \cite{Ofcom2020}, for example, the inclusion of news and institutional accounts may act as a proxy for public attention and opinion at large.

With regard to bots: previous large-scale studies of Twitter data have demonstrated the influence bots can have on the exposure of human accounts to emotional content \cite{Stella2018} and the extent to which they can distort the discussion on certain topics \cite{Bessi_Ferrara_2016}. More recently and in the context of the current pandemic, bots have been shown to have a significant role in promoting political conspiracy theories \cite{ferrara2020covid}. By ignoring retweets and using unique original tweets only, we mitigate to some extent the potential effect of bots, which have previously been shown to engage in retweeting behaviour significantly more frequently than they do the creation of original content \cite{Bessi_Ferrara_2016}. It is nonetheless likely that, even if the hypothesised psychophysical numbing effect is genuine, our observations are partly attributable to the nature of content generated by these non-human accounts. 

Furthermore, we} stress that the results presented in this paper may be indicative only of the responses of Twitter users posting from each of these countries in each of these languages, so extrapolating these results to the broader population will only be possible with a better understanding of the biases present in, and representativeness of, the dataset at hand. { While the demography of Twitter users has been to some extent mapped for the United States (see e.g. \cite{Greenwood2016}) and the United Kingdom (see e.g. \cite{Sloan2017}), it is difficult to find similar studies for the remaining countries in our dataset, and thus to interpret these country-level differences in terms of potentially differing demographic representation on Twitter. We nonetheless advance this as a factor that possibly contributes to our results. 

We also reiterate that our analysis has been crude in that we make use of a single dictionary for each language when extracting linguistic features from our data. This ignores important differences in dialect and language use between different nationalities and cultures, and can result in the systematic omission of certain linguistic features \cite{rudra2016understanding, Mathur2019} which may also contribute to the observed differences between countries. Further important differences between countries which may help to account for the observed results are differences in the importance of religion in each of the considered countries. The set of countries under consideration here span the full spectrum of importance assigned to religion \cite{Hackett2018}, and attitudes towards death and the framing of mortality may vary accordingly by country. Despite these difficulties inherent to the empirical analysis of social media data, we nonetheless hope that our work inspires further investigations into the use of natural language processing and cognitive network science to investigate the prevalence of psychophysical numbing in naturalistic contexts.}

%%% \cite{Sloan2017} says UK Twitter users are younger than UK demographics and more likely to be managerial, administrative, professional occupations. 
%%% \cite{Greenwood2016} says US Twitter users are also younger than US demographics, and more likely to be college-educated

\section{Declaration}

\subsection{Acknowledgements}
The authors would like to thank Mirta Galesic, Rodrigo Leal Cervantes, Rita Maria del Rio Chanona, Fran\c{c}ois Lafond, and J. Doyne Farmer for helpful feedback, and to the Oxford INET Complexity Economics group for stimulating discussions. The authors also thank the anonymous reviewers whose suggestions significantly improved the quality of the final paper. 

\subsection{Funding}
BK acknowledges funding from the Conacyt-SENER: Sustentabilidad Energética scholarship and JD acknowledges funding from the EPSRC Industrially Focused Mathematical Modelling Centre for Doctoral Training. 

\subsection{Competing interests}

The authors declare that they have no competing interests.

\subsection{Availability of data and materials}

The Twitter data used in the manuscript is collected and maintained by Banda {\it et al.} at the Panacea Lab \cite{banda_juan_m_2020_3757272}, and it is available at their website \url{http://www.panacealab.org/covid19}. The data on Covid-19 confirmed cases and deaths were obtained from the ``Coronavirus Pandemic (COVID-19)'' page of the Our World in Data website \cite{owidcoronavirus}, and the stable URL for this data is \url{https://covid.ourworldindata.org/data/owid-covid-data.csv}.

% Release of our data?

\subsection{Authors' contributions}

BK and JD both conceived the idea, carried out the analysis, and wrote, read, and approved the final manuscript.

\subsection{Abbreviations}

\begin{itemize}
    \item LIWC: Linguistic Inquiry and Word Count.
    \item WHO: World Health Organization.
    \item NLS: National linguistic score.
\end{itemize}

%%%%%%%%%%%%%%%%%%%%%%%%%%%%%%%%%%%%%%%%%%%%%%%%%%%%%%%%%%%%%%%%%%%%%%%%%%%%%%%%%%%%%
%%%% This is the Bibliography where you will cite your sources used in the paper %%%%

\bibliographystyle{ieeetr}
\bibliography{references}

\newpage
\begin{appendices}
\section{Further model comparison}\label{app:comparison}
In this section, we present further results of our models to give a more complete overview of their quality. Besides the Weber-Fechner law and power law models (see Eqs. (\ref{eq:WFL}) and (\ref{eq:powerlaw})), we use the following linear relationship between $s$ and $p$ as our benchmark model
\begin{equation}
    p(t) = a\cdot s(t) + b,
    \label{eq:linear}
\end{equation}
where $a$ and $b$ are parameters. We summarize our results for the linear model in Table \ref{tab:linear}.

For all models, we compute the $R^2$ values
\begin{equation}
    R^2 = 1 - \frac{ \sum_{t=1}^n e(t)^2  }{ (n-1) \sigma_p^2 },
    \label{eq:rsquared}
\end{equation}
where $ e(t) = p(t) - \hat{p}(t) $ is the model residual, $\sigma_p^2 = \sum_{t=1}^n ( p(t) - \mu_p )^2 / (n-1) $ is the variance of $p(t)$, and $n$ is the sample size. The $R^2$ values for all models are summarized in Table \ref{tab:R2}. (Note that as the power law model implies a log-normal residual, the $R^2$ values can be negative.) From this table we see that, once again, the Weber-Fechner law is generally a better fit to the data across all countries, but that the power law and Weber-Fechner models are often comparable and significantly better than the linear model.

\begin{figure}
    \centering
    \includegraphics[width=1\linewidth]{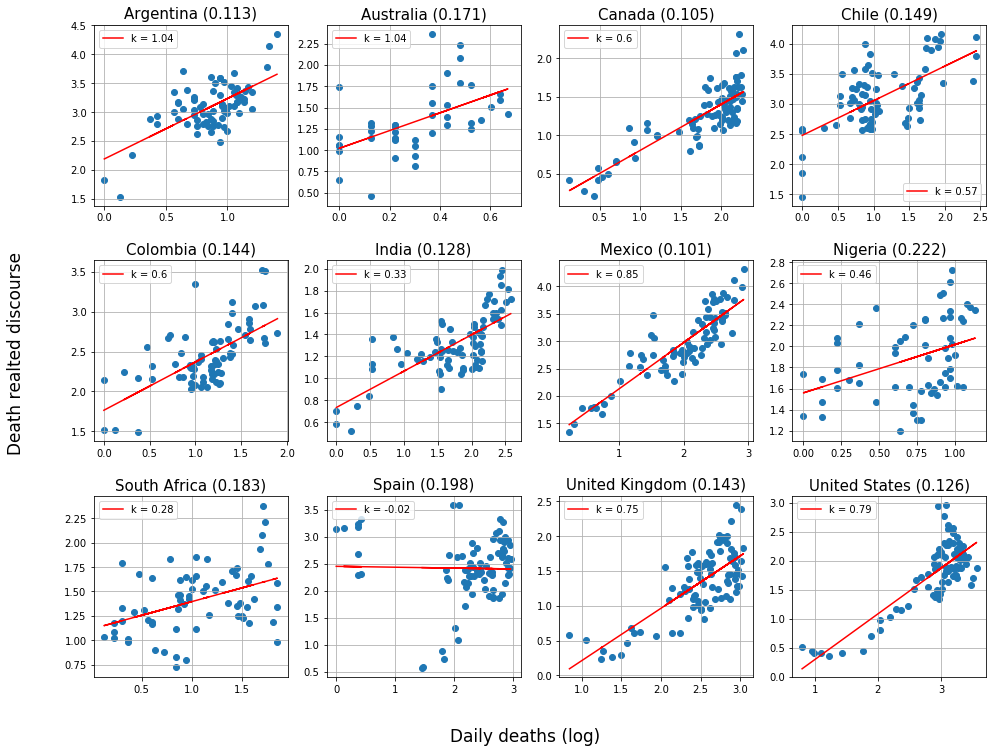}
    \caption{Resulting scatter plot for the \textbf{Weber-Fechner law} model fit, where each panel shows a different country with their corresponding  NRMSE in parenthesis (the lower the better). }
    \label{fig:scatterplot_weber}
\end{figure}

\begin{figure}
    \centering
    \includegraphics[width=1\linewidth]{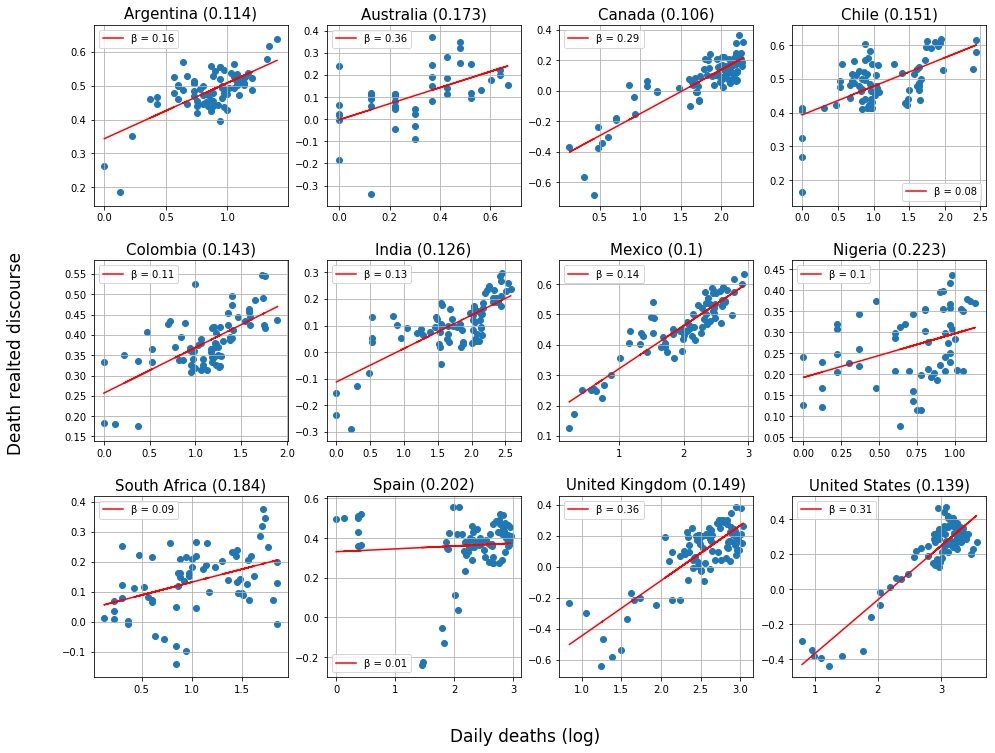}
    \caption{Resulting scatter plot for the \textbf{power law} model fit, where each panel shows a different country with their corresponding  NRMSE in parenthesis (the lower the better).}
    \label{fig:scatterplot_powerlaw}
\end{figure}

We also show in Figures \ref{fig:scatterplot_weber} and \ref{fig:scatterplot_powerlaw} scatterplots of the Death NLSs against the logarithm of the daily number of deaths in each country, with the $y$-axis in linear- and log-scales, respectively. Red lines indicate the line of best fit, with the slope equal to $k$ and $\beta$ in Eqs. \ref{eq:WFL} and \ref{eq:powerlaw}, respectively.

%TABLE
\begin{table}[h]
\centering
\resizebox{\textwidth}{!}{\begin{tabular}{l|cc||ccc||cccc}
\toprule
{}               & $a$ & $b$ & 95\% CI &       $t$ &  $P>|t|$ & $R^2$ &    NRMSE & $n$ \\ %  AIC &
\textbf{Country} &       &           & ($a$) & ($a$)   & ($a$)  &       &              &     \\ %    &
\midrule
\textbf{Argentina     } &  0.057 &     2.603 &   0.04 -- 0.074 &   6.79 &     0.0 &  0.387 &   0.116 &  75 \\ %  49.2 & 
\textbf{Australia     } &  0.192 &     0.912 &  0.087 -- 0.298 &   3.67 &  0.0007 &  0.247 &   0.175 &  43 \\ %  31.4 & 
\textbf{Canada        } &  0.005 &     0.759 &  0.005 -- 0.006 &  11.46 &     0.0 &  0.607 &   0.117 &  87 \\ % 6.8 & 
\textbf{Chile         } &  0.005 &     2.969 &  0.003 -- 0.007 &   4.58 &     0.0 &  0.216 &    0.17 &  78 \\ % 103.8 & 
\textbf{Colombia      } &  0.016 &     2.147 &   0.011 -- 0.02 &   7.11 &     0.0 &  0.409 &   0.145 &  75 \\ %  33.2 & 
\textbf{India         } &  0.002 &     1.074 &  0.002 -- 0.003 &  10.14 &     0.0 &  0.566 &   0.125 &  81 \\ % -40.5 & 
\textbf{Mexico        } &  0.003 &     2.428 &  0.002 -- 0.003 &  10.97 &     0.0 &  0.613 &   0.133 &  78 \\ %  81.3 & 
\textbf{Nigeria       } &  0.047 &     1.589 &  0.021 -- 0.073 &   3.61 &  0.0006 &  0.184 &   0.218 &  60 \\ %  42.5 & 
\textbf{South Africa  } &  0.006 &     1.295 &   0.002 -- 0.01 &   2.95 &  0.0045 &  0.119 &   0.188 &  66 \\ %  37.5 & 
\textbf{Spain         } &      0 &     2.248 &   -0.0 -- 0.001 &   1.98 &  0.0506 &  0.047 &   0.193 &  82 \\ %  148.4 & 
\textbf{United Kingdom} &  0.001 &     0.877 &  0.001 -- 0.001 &   7.69 &     0.0 &  0.399 &   0.166 &  91 \\ %  80.4 & 
\textbf{United States } &      0 &     1.235 &    0.0 -- 0.001 &   6.84 &     0.0 &  0.352 &   0.179 &  88 \\ %  119.1 & 
\bottomrule
\end{tabular}}
\caption{ Results for the linear model defined in Eq. (\ref{eq:linear}). }
\label{tab:linear}
\end{table}

%TABLE
\begin{table}%[H]
\centering
% \resizebox{\textwidth}{!}
{\begin{tabular}{l|ccc}
\toprule
{$R^2$} & Power law & Weber-Fechner law & Linear relationship \\
\textbf{Country       } &           &             &                     \\
\midrule
\textbf{Argentina     } &     0.411 &       \textbf{0.421} &               0.387 \\
\textbf{Australia     } &     0.259 &       \textbf{0.275} &               0.247 \\
\textbf{Canada        } &     0.678 &       \textbf{0.683} &               0.607 \\
\textbf{Chile         } &     0.382 &       \textbf{0.395} &               0.216 \\
\textbf{Colombia      } &     \textbf{0.425} &       0.319 &                0.28 \\
\textbf{India         } &     \textbf{0.558} &       0.397 &               0.477 \\
\textbf{Mexico        } &      \textbf{0.78} &       0.775 &               0.613 \\
\textbf{Nigeria       } &    \textbf{0.143} &       0.004 &               0.007 \\
\textbf{South Africa  } &      0.16 &       \textbf{0.171} &               0.119 \\
\textbf{Spain         } &    -0.042 &           0 &               \textbf{0.047} \\
\textbf{United Kingdom} &     0.514 &       \textbf{0.555} &               0.399 \\
\textbf{United States } &     \textbf{0.608} &       0.556 &                0.24 \\ \hline
\textbf{Mean          } &      0.36 &       \textbf{0.379} &               0.303 \\
\textbf{Proportion of best fits   } &   41.7 \% &     \textbf{50} \% &               8.33 \% \\
\textbf{Proportion of second-best fits   } &   \textbf{66.7} \% &     33.3 \% &               0 \% \\
\bottomrule
\end{tabular}}
\caption{Comparison of $R^2$ between the power law model of Eq. (\ref{eq:powerlaw}), the Weber-Fechner model of Eq. (\ref{eq:WFL}) and a linear relationship between variables, which we use as a benchmark model. Higher values indicate better models.}
\label{tab:R2}
\end{table}

%SOME SCATTERPLOTS
%FIGURE

%FIGURE

\newpage

{ \section{National Linguistic Scores}\label{app:peaks}

{  \subsection{Exogenous peaks in the National Linguistic Scores}

In this section, we address significant deviations in the National Linguistic Scores from our proposal of psychophysical numbing as an explanation for their trends over the observation period, and suggest possible explanations for their occurrence, see Table \ref{table:exo_peaks}. We stress that the following Table might be prone to error although we double checked every peak.

%TABLE
\begin{table}%[h!]
\centering
% \resizebox{\textwidth}{!}{%
\begin{tabularx}{\textwidth}{l|c|c|X}
\toprule
\textbf{Country} & \textbf{Date} & \textbf{Category} & \textbf{Description} \\
\midrule
United States & June 2, 2020 & Anger & Donald Trump responding near the White House to the protests against the murder of George Floyd \cite{stepehen_collinson_2020}  \\
\makecell{Australia, Nigeria, 
\\ United Kingdom} & June 6-7, 2020 & Anger & Worldwide protests against the murder of George Floyd in the United States \cite{cnn_world_2020} \\
Chile & May 13-15, 2020 & Anger & Chilean Health Ministry announces total lockdown \cite{health_ministry_2020}.  \\
Argentina & March 20, 2020 & Death & President Alberto Fernandez announces total national lockdown \cite{elpais_2020}. \\
Australia & May 20-26, 2020 & Death & New Covid-19 deaths after several days without casualties. Moreover, the murder of George Floyd in the United States took place on May 25 \cite{nbc_2020}.  \\
Canada & May 2, 2020 & Death & \textit{Unknown} \\
Colombia & March 22, 2020 & Death & Shootout in a prison triggered by prisoners demanding better hygiene conditions for Covid-19 results in 23 deaths. \cite{efe_2020} \\
India & March 29, 2020 & Death & \textit{Unknown} \\
Nigeria & April 18, 2020 & Death & Many important African political figures die from Covid-19 this day \cite{AfricanFigures}. \\
Spain & May 25, 2020 & Death & Correction in Covid-19 data repositories show negative daily deaths \cite{diario_2020}. \\
Spain & June 1, 2020 & Death & First day in Spain without Covid-19 deaths \cite{financiero_2020}. \\
United Kingdom & April 10, 2020 & Death & The United Kingdom surpasses 10,000 Covid-19 deaths \cite{bbc_2020}. \\
United States & May 25, 2020 & Death & Murder of George Floyd \cite{nbc_2020} \\

\bottomrule
\end{tabularx}%
% }
\caption{ { A list of plausible explanations for anomalous peaks observed in Figure \ref{fig:liwc_panel}. Most of these peaks arise from the murder of George Floyd and the consequent protests. We focused mainly on the Death and emotionally-charged categories as these are the ones that are most related to the psychophysical numbing effect we describe in the main text.} }
\label{table:exo_peaks}
\end{table}
}
}

\section{Word co-occurrence analysis}\label{CoApp}

\subsection{Further technical details on co-occurrence network construction}\label{DetailsCo}

In constructing the word co-occurrence networks presented in Section \ref{PsychNumb}, we preform basic text preprocessing, including taking the lower-case form of all letters, removing URLs, removing punctuation, and removing the following small set of stopwords from the vocabulary:
\begin{center}
    \texttt{to, today, too, has, have, like}.
\end{center}
We retain hashtags, since LIWC also recognises hashtags and because hashtags are an essential aspect to communications on Twitter. It is also necessary to account for the fact that a number of ``words'' appearing in the LIWC dictionary are in fact regular expressions to which many complete words in the Twitter dataset map. For example, the ``word'' ``isolat*'' appears in the English LIWC dictionary, to which each of the following words would map: ``isolate'', ``isolated'', ``isolating''. Thus, construction of the word co-occurrence networks $G'_i$ involves a two-step procedure: first, constructing the raw word co-occurrence networks $G_i$, in which the nodes are words exactly as they appear in the Twitter dataset; and then reducing this to a quotient graph $G'_i$ by contracting nodes in $G_i$ that are matched by the same regular expression in the LIWC dictionary. More formally: the LIWC dictionary implies an equivalence relation $\sim$ on the vocabulary $\mathcal{V}$ implied by the Twitter dataset, such that $v\sim u$ for words $v, u \in \mathcal{V}$ if both $v$ and $u$ are matched by the same regular expression in the LIWC dictionary. The weights of edges between nodes $v' \subset \mathcal{V}$ and $u' \subset \mathcal{V}$ in $G'_i$ are then taken to be
\begin{equation}
    w_{G'_i}(u', v') = \sum_{u\in u', v\in v'} w_{G_i}(u, v),
\end{equation}
where $w_{G}(x, y)$ is the weight of edge $(x, y)$ in $G$. Note that $w_{G}(x, y) = w_{G}(y, x)$ and $w_{G}(x, y) = 0$ if $(x, y)$ is not an edge in $G$.

{  To construct the higher-frequency sequences of snapshots, we impose a minimum document frequency of $5\times 10^{-3}$ ($2.5\times 10^{-3}$ for Spanish tweets) for each term in the vocabulary in order to reduce the effect of noise. In Table \ref{tab:Quant_WordCo}, we summarise the approximate number $N_{\rm tweets}$ of tweets per snapshot for each country. The number of tweets per snapshot for each country was chosen in order that each country had approximately the same number of data points separated by approximately 3 days, and such that edge effects did not yield a final snapshot with a disproportionately low number of tweets. While this ultimately led to some snapshots representing aggregation over longer periods than others, this yielded sequences of networks that are comparable in terms of their total strength and order, enabling reasonably fair comparison of the modularities of the partition induced by the Death and Affect LIWC categories.

\begin{table}[]
\begin{center}
\begin{tabular}{lcccc}
    \specialrule{.1em}{.05em}{.05em}
{\bf Country} & {\bf Argentina} & {\bf Australia} & {\bf Canada} & {\bf Chile} \\
  %  \hline
${N_{\rm tweets}}$ & $3.5\times 10^4$  & $2\times 10^{4}$ & $5 \times 10^{4}$ & $1\times 10^4$ \\
    \specialrule{.1em}{.05em}{.05em}
{\bf Country} & {\bf Colombia} & {\bf India} & {\bf Mexico} & {\bf Nigeria} \\
  %  \hline
$N_{\rm tweets}$ & $1.5\times 10^4$ & $5\times 10^{4}$ & $4 \times 10^4$ & $2 \times 10^{4}$\\
    \specialrule{.1em}{.05em}{.05em}
{\bf Country} & {\bf South Africa} & {\bf Spain} & {\bf United Kingdom} & {\bf United States} \\
  %  \hline
${N_{\rm tweets}}$ & $1\times 10^4$ & $5 \times 10^4$ & $1 \times 10^{5}$ & $1\times 10^5$\\
    \specialrule{.1em}{.05em}{.05em}
\end{tabular}
\caption{{ The number of tweets taken per snapshot for each country.}}\label{tab:Quant_WordCo}
\end{center}
\end{table}

}

\subsection{Word co-occurrence networks for Spanish-language tweets}\label{SpCo}

For completeness, we provide here the word co-occurrence graphs for the Spanish language tweets. We omit a discussion of the results, since similar conclusions can be drawn from these as in the English counterparts.

%FIGURE!
\begin{figure}[h!]
\centering
	\begin{subfigure}[b]{0.49\linewidth}
	\includegraphics[width=\textwidth]{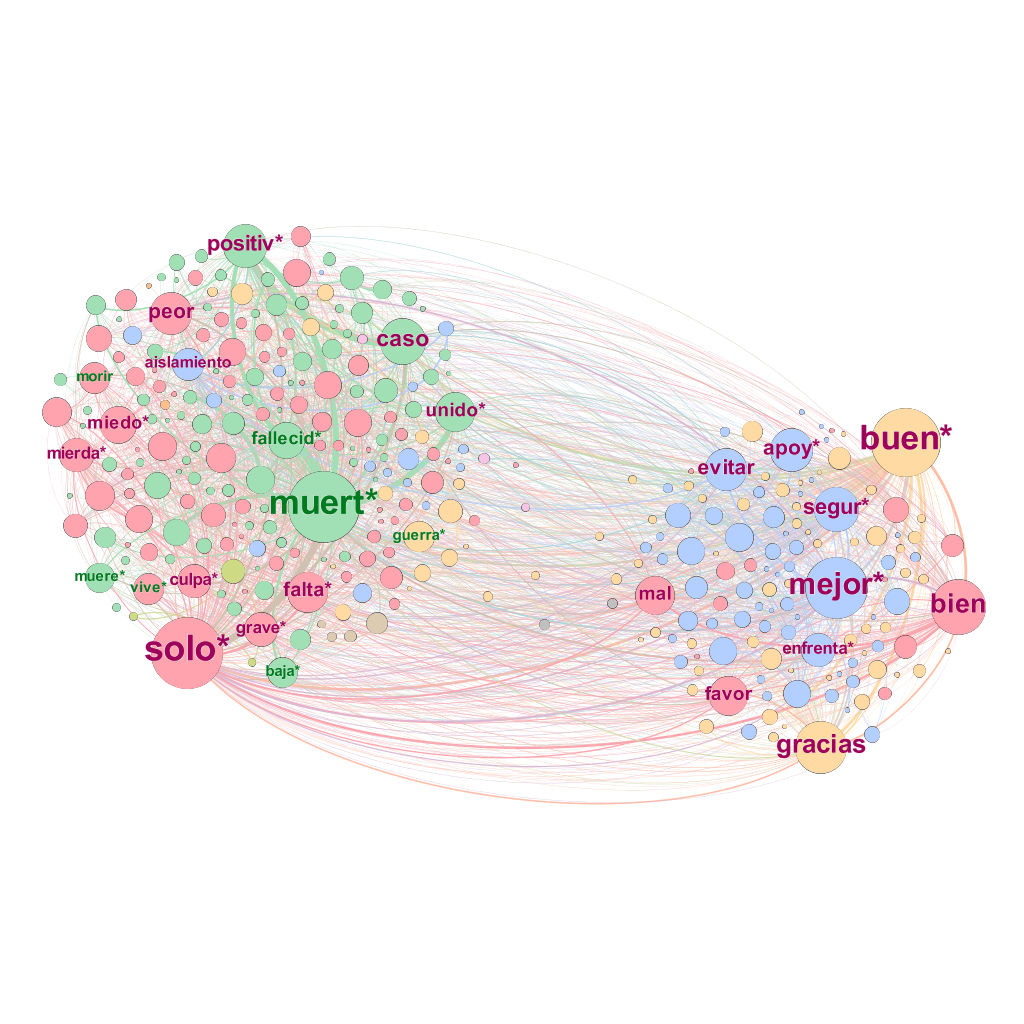}
	\caption{March 11th to April 9th, 2020. }	
	\end{subfigure}
	\begin{subfigure}[b]{0.49\linewidth}	
	\includegraphics[width=\textwidth]{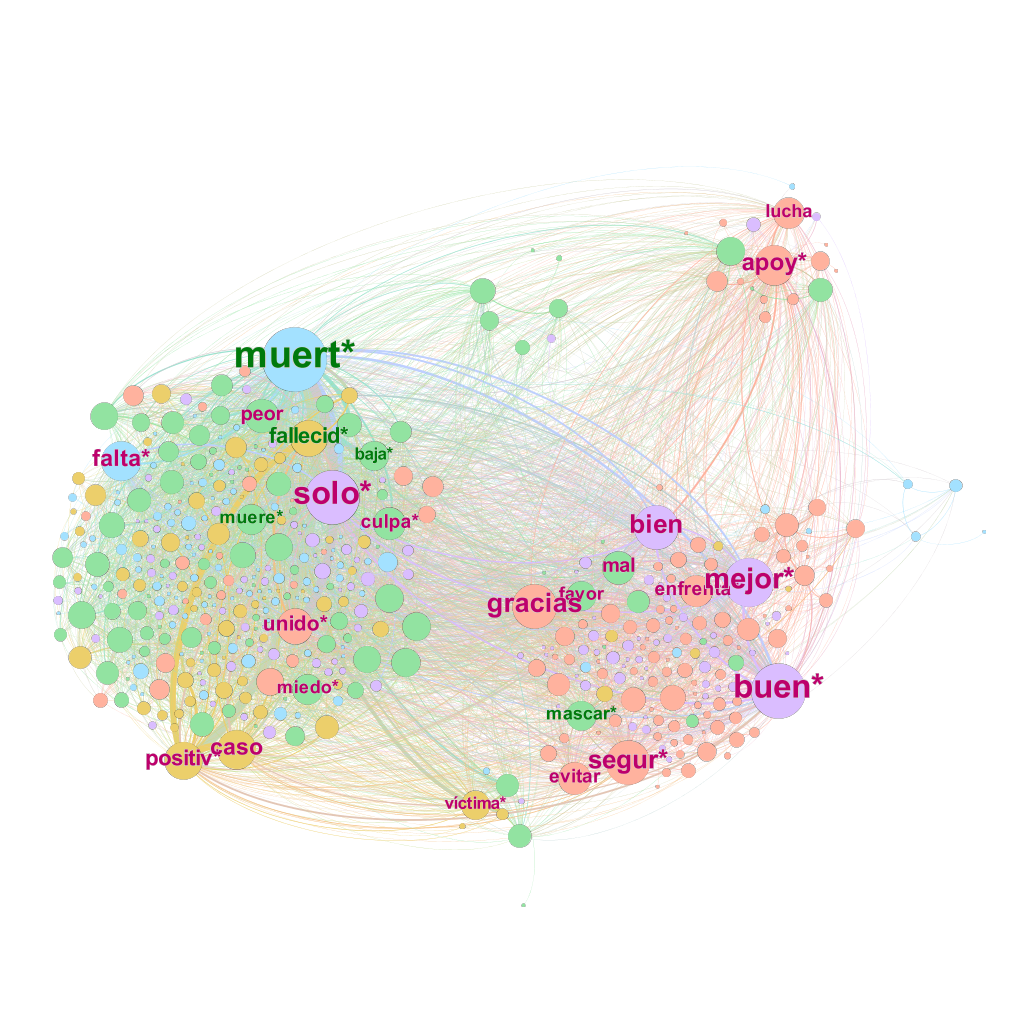}
	\caption{April 10th to May 23rd, 2020.}	
	\end{subfigure}
	\begin{subfigure}[b]{0.49\linewidth}	
	\includegraphics[width=\textwidth]{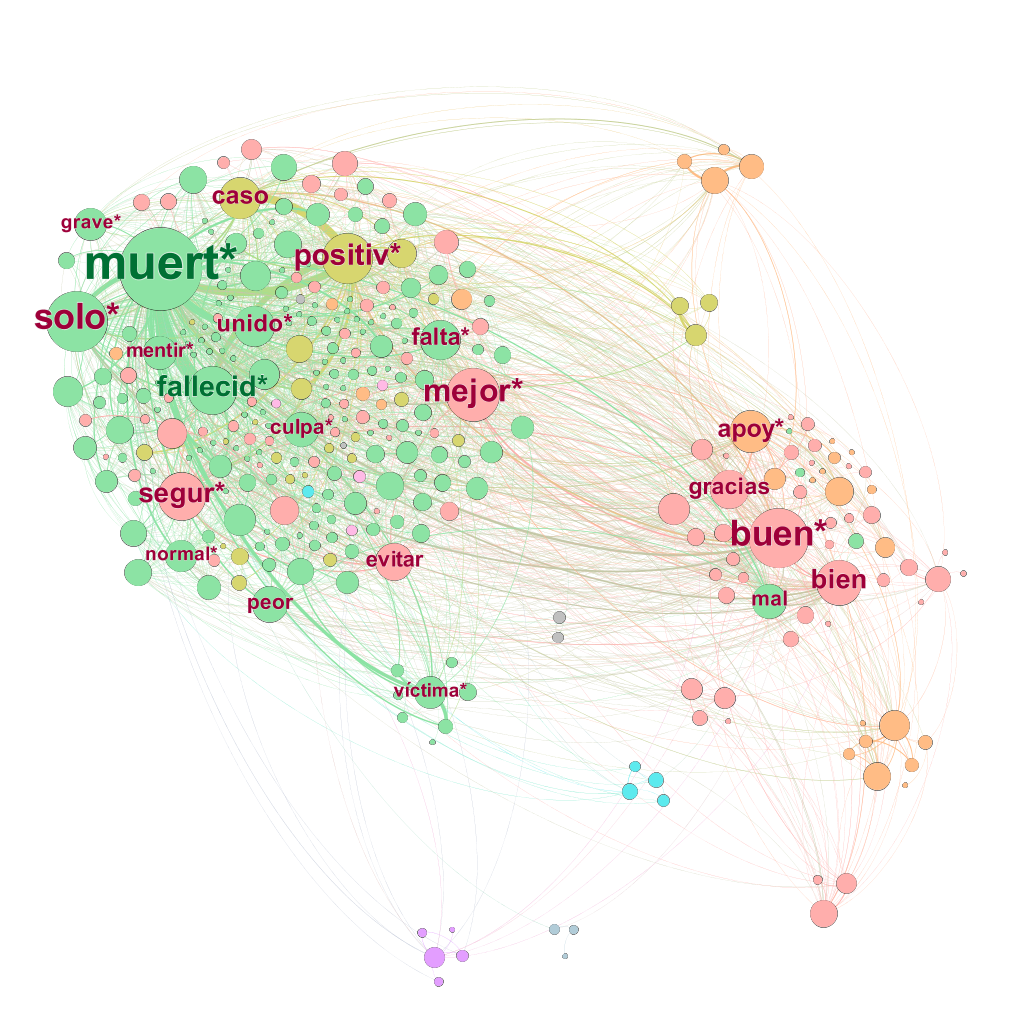}
	\caption{May 24th to June 14th, 2020.}	
	\end{subfigure}
	\caption{Snapshots of the word co-occurrences associated with death (``muerte'', green labels) and affect (``afecto'', red labels) for Spanish-language tweets aggregated across all analyzed countries in three different time windows (see sub-captions). The nodes are coloured based on the community labels obtained by maximising modularity using the Louvain algorithm \cite{Blondel_2008}. We filtered edges with weight below $20$ co-occurrences for visualisation purposes.}
\end{figure}

\section{Covid-19 epidemiological data}
\begin{figure}[h]
    \centering
    \includegraphics[width=0.7\linewidth]{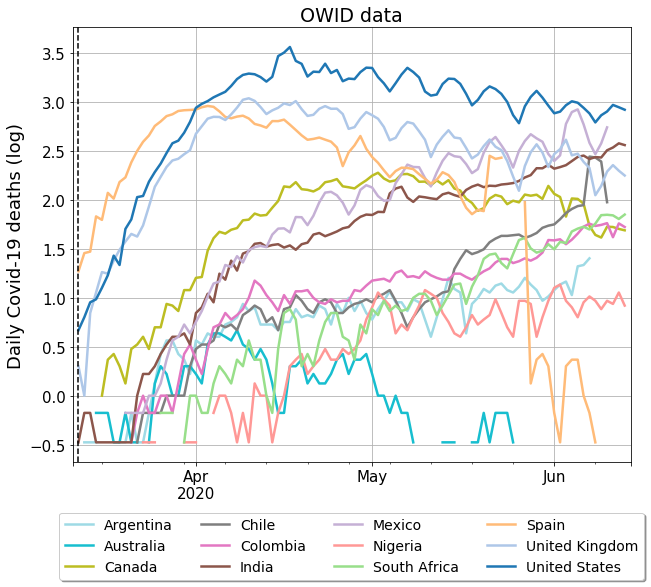}
    \caption{
    Daily deaths related to Covid-19 for each of the countries in our analysis (see legend) from March 11 to June 14, 2020. 
    }
    \label{fig:covid_data}
\end{figure}

We include this section as a reference for the actual number of deaths in each country for the period we analysed throughout the paper, which we present in Fig. \ref{fig:covid_data}.

\end{appendices}

\end{document}